\documentclass[aps,prd, twocolumn, showpacs, nofootinbib]{revtex4-1}
\usepackage{bm}
\usepackage{amsmath}
\usepackage{amsfonts}          
\usepackage{latexsym}
\usepackage{amssymb}
\usepackage{amsmath}
\usepackage{dsfont}
\usepackage{url}
\usepackage{graphicx}
\usepackage[applemac]{inputenc}  
\usepackage[english]{babel}
\usepackage{csvsimple}
\usepackage{color}

\numberwithin{equation}{section}
\renewcommand{\theequation}{\arabic{section}.\arabic{equation}}

\newcommand{\be}{\begin{equation}}
\newcommand{\ee}{\end{equation}}

\newcommand{\ba}{\begin{align}}
\newcommand{\ea}{\end{align}}

\newcommand{\lb}{\left(}
\newcommand{\rb}{\right)}

\newcommand{\av}{\bm{a}}
\newcommand{\na}{(\bm{n}\cdot\bm{a})}
\newcommand{\pa}{(\bm{p} \cdot \bm{a})}
\newcommand{\enpa}{\left( (\bm{n} \times \bm{p})\cdot \bm{a} \right)}
\newcommand{\enpcz}{\left( (\bm{n} \times \bm{p})\cdot \bm{\chi}_0 \right)}

\newcommand{\az}{\bm{a}_0}
\newcommand{\naz}{(\bm{n}\cdot\bm{a}_0)}

\newcommand{\enpaz}{\left( (\bm{n} \times \bm{p})\cdot \bm{a}_0 \right)}

\newcommand{\p}{\bm{p}}
\newcommand{\cz}{\bm{\chi}_0}
\newcommand{\cu}{\bm{\chi}_1}
\newcommand{\cd}{\bm{\chi}_2}
\newcommand{\np}{(\bm{n}\cdot \bm{p})}
\newcommand{\ncz}{(\bm{n}\cdot \bm{\chi}_0)}
\newcommand{\ncu}{(\bm{n}\cdot \bm{\chi}_1)}
\newcommand{\ncd}{(\bm{n}\cdot \bm{\chi}_2)}

\newcommand{\pcu}{(\bm{p}\cdot \bm{\chi}_1)}
\newcommand{\pcd}{(\bm{p}\cdot \bm{\chi}_2)}
\newcommand{\cucd}{(\bm{\chi}_1 \cdot \bm{\chi}_2)}

\newcommand{\cicj}{(\bm{\chi}_i \cdot \bm{\chi}_j)}
\newcommand{\nci}{(\bm{n}\cdot \bm{\chi}_i)}
\newcommand{\ncj}{(\bm{n}\cdot \bm{\chi}_j)}
\newcommand{\pci}{(\bm{p}\cdot \bm{\chi}_i)}
\newcommand{\pcj}{(\bm{p}\cdot \bm{\chi}_j)}

\newcommand{\nai}{(\bm{n}\cdot \bm{a}_i)}

\newcommand{\pai}{(\bm{p}\cdot \bm{a}_i)}
\newcommand{\paj}{(\bm{p}\cdot \bm{a}_j)}

\newcommand{\mud}{\frac{m_1}{m_2}}
\newcommand{\mdu}{\frac{m_2}{m_1}}

\begin{document}

\title{A new effective-one-body Hamiltonian with next-to-leading order spin-spin coupling} 
\author{Simone Balmelli}
\affiliation{Physik-Institut, Universit\"at Z\"urich, Winterthurerstrasse 190, 8057 Z\"urich, Switzerland}

\author{Thibault Damour}
\affiliation{Institut des Hautes \'Etudes Scientifiques, 91440 Bures-sur-Yvette, France}
\begin{abstract}
We present a new effective-one-body (EOB) Hamiltonian with next-to-leading order (NLO) spin-spin coupling for black hole binaries endowed with arbitrarily oriented spins.
The Hamiltonian is based on the model for parallel spins and equatorial orbits developed in [Physical Review D \textbf{90}, 044018 (2014)], but differs from it in several ways.
In particular, the NLO spin-spin coupling is not incorporated by a redefinition of the centrifugal radius $r_c$, but by separately modifying certain sectors of the Hamiltonian, which are identified according to their dependence on the momentum vector.
The gauge-fixing procedure we follow allows us to reduce the 25 different terms of the NLO spin-spin Hamiltonian in Arnowitt-Deser-Misner coordinates to only 9 EOB terms.
This is an improvement with respect to the EOB model recently proposed in [Physical Review D \textbf{91}, 064011 (2015)], where 12 EOB terms were involved. 
Another important advantage is the remarkably simple momentum structure of the spin-spin terms in the effective Hamiltonian, which is simply quadratic up to an overall square root.
Moreover, a Damour-Jaranowski-Sch\"afer-type gauge could be established, thus allowing one to concentrate, in the case of circular and equatorial orbits, the whole spin-spin interaction in a single radial potential.
\end{abstract}


\maketitle

\section{Introduction}
The increasing interest in the modeling of gravitational waveforms from coalescing binaries, strongly motivated by the construction of ground-based detectors such as Virgo \cite{virgo} or the now operating advanced LIGO \cite{TheLIGOScientific:2014jea} instruments, has led in the last decade to a significant effort in calculating spin effects in the post-Newtonian (PN) two-body problem beyond the leading order (LO). 
The spin-orbit coupling at the next-to-leading-order (NLO) was first derived in harmonic coordinates \cite{Faye:2006gx, Blanchet:2006gy}, and then within an Arnowitt-Deser-Misner (ADM) formalism \cite{Damour:2007nc}. 
The ADM approach (see especially the formalism developed in Ref.~\cite{Steinhoff:2008zr}) has been quite fruitful, since it has also allowed the calculation of the next-to-next-to-leading order (NNLO) spin-orbit coupling \cite{Steinhoff:2009ei,Hartung:2011te} and of the NLO spin-spin\footnote{In this paper, ``spin-spin'' refers to any interaction quadratic in the spins, i.e., $\propto$ $S_1^2$, $S_2^2$ and $S_1S_2$.} coupling \cite{Hergt:2008jn, Steinhoff:2008ji, Steinhoff:2007mb}.
A method based on Effective Field Theory techniques \cite{Goldberger:2004jt} has also been able to derive the same results (see e.g. \cite{Levi:2015msa}), and is expected to complete soon the (full, physically relevant) spin-spin coupling at the NNLO accuracy \cite{Levi:2015ixa}. 

Past work has shown that the most efficient way of using PN-expanded results to describe the dynamics of coalescing binaries is to encode them into an effective-one-body (EOB) model \cite{Buonanno:1998gg,Buonanno:2000ef,Damour:2000we,Damour:2001tu,Damour:2008gu}. 
This objective has been pursued in different versions of the EOB \cite{Damour:2001tu,Buonanno:2005xu,Damour:2008qf, Barausse:2009aa,Barausse:2009xi,Pan:2010hz, Nagar:2011fx, Barausse:2011ys,Pan:2013rra} for both the spin-orbit coupling (up to NNLO) and the spin-spin coupling (up to LO).
 
More recently, an EOB Hamiltonian reproducing the correct NLO spin-spin coupling has been proposed \cite{Balmelli:2013zna, Balmelli:2013err, Balmelli:2015lva}, where the terms in question are included by a subleading-order modification of various squared-spin terms.
An unpleasant feature of this approach is that the so-obtained effective squared-spin acquires a momentum dependence that cannot be removed by any gauge tuning, and that greatly complicates the analytic form of the Hamiltonian.
In addition, the momentum-dependent terms in question are non-zero even in the most simple case of circular and equatorial orbits, which prevents one from having a direct insight into the dynamics by means of a radial potential $A$, as is the case for the models with just LO spin-spin coupling (see e.g. \cite{Damour:2001tu,Buonanno:2005xu,Damour:2008qf, Nagar:2011fx}).

Recently, Ref.~\cite{Damour:2014sva} has proposed a new EOB description of binary black holes with parallel spins, moving along equatorial orbits.
The EOB Hamiltonian of Ref.~\cite{Damour:2014sva} incorporates a reformulation of the NLO spin-spin terms of Ref.~\cite{Balmelli:2013zna}, but presents some basic structural differences with respect to Refs.~\cite{Damour:2001tu, Damour:2008qf,Balmelli:2013zna, Balmelli:2015lva}.
The most important ones are the introduction of a new variable (the centrifugal radius $r_c$), which plays a central role for the description of quadratic spin effects, and a simplification of the spin-orbit structure.

The present work is meant as an improvement of both Ref.~\cite{Balmelli:2015lva} and Ref.~\cite{Damour:2014sva}.
It will overcome the problematic features of Ref.~\cite{Balmelli:2015lva} discussed above, while staying as close as possible to the new formalism and ideas introduced in Ref.~\cite{Damour:2014sva}.
Our final result will be an EOB Hamiltonian describing arbitrarily oriented spinning black holes whose structure is physically transparent and quite close to that of the Hamiltonian describing the dynamics of a test-particle in a Kerr background.
As a bonus, our Hamiltonian will make manifest six hidden symmetries of the NLO spin-spin coupling, thereby allowing one to describe the latter coupling by means of only 9 terms (instead of the 25 terms present in their ADM formulation).

In Sec~\ref{sec:NLOss}, which is the core of the paper, our whole procedure is sequentially presented until the main results are obtained; in particular, Sec~\ref{sec:Kerr} revisits the Kerr Hamiltonian and develops, from this limiting case, the basic ideas to be applied in the EOB case; Sec.~\ref{sec:Horbeff} introduces the EOB model from which we start, and Sec~\ref{sec:Can} defines the transformation between the ADM and EOB coordinates; Sec.~\ref{sec:Gauge} discusses two possible gauge choices, eventually opting for a single one, which leads to an identification of some forms quadratic in the spins that must be inserted into the EOB model to reproduce the NLO spin-spin coupling; Sec.~\ref{sec:resum} proposes a resummation of the results into a final EOB Hamiltonian;
Sec.~\ref{sec:quad} provides a more detailed description of the quadratic forms, with some details about their eigenvalue decomposition and their positivity properties.
In Sec~\ref{sec:LSO}, the spin-orbit sector is discussed with some emphasis about the resummation choices  of the gyro-gravitomagnetic factors.
The physical characteristics of the last stable orbit (LSO) for equal masses and equal, aligned spins, are then computed and compared with the predictions of other EOB models.
Finally, the Appendix briefly discusses some unexpected ``symmetries''  in the coefficients of the quadratic forms.
Througout the paper we use geometrical units with $G \equiv c \equiv 1$.

\section{A new effective-one-body description of the next-to-leading order spin-spin coupling}
\label{sec:NLOss}

Let us recall that one of the basic features of the EOB formalism is to represent the Hamiltonian of a (comparable-mass and comparable-spin) two-body system in the form

\be
\label{eq:Heob}
H_\text{EOB} = M\sqrt{1 + 2\nu\lb \frac{H^\text{eff}}{\mu} -1 \rb},
\ee
where the ``effective'' Hamiltonian $H^\text{eff}$ is a \emph{deformed version} of the Hamiltonian describing the dynamics of a (spinning) test-particle in a Kerr background.
The EOB effective Hamiltonian is decomposed as

\be
\label{eq:orb+so}
H^\text{eff} = H_\text{orb} + H_\text{so},
\ee
where the spin-orbit part $H_\text{so}$ gathers the contributions that are odd in the spins (i.e. linear, cubic, etc.), while the orbital part $H_\text{orb}$ those that are even in the spins (i.e. spin-independent, and then quadratic, quartic, etc.).

\subsection{Structure of the Kerr Hamiltonian in Cartesian-like coordinates}
\label{sec:Kerr}

As an orientation towards defining a new EOB Hamiltonian incorporating NLO spin-quadratic effects, let us reexamine the structure of the limiting case (to which $H^\text{eff}$ should reduce in the extreme mass ratio limit) of the Hamiltonian of a (non spinning) test-particle in a Kerr background. 
For this Kerr dynamics, and for the special case of equatorial orbits, Ref.~\cite{Damour:2014sva} has highlighted the role played by the \emph{centrifugal} radius 

\be
\label{eq:rcKerr}
r_c = \sqrt{r^2 + a^2 + \frac{2Ma^2}{r}},
\ee
where $r$ is the Boyer-Lindquist radial coordinate.
In Eq.~(\ref{eq:rcKerr}), $M$ denotes the mass of the considered Kerr black hole, and $a$ its Kerr parameter.
The \emph{orbital} sector of the test-particle Kerr Hamiltonian (after setting apart, similarly to Eq.~(\ref{eq:orb+so}), its spin-orbit sector) takes the form (in polar coordinates)

\be
\label{eq:HorbKerr}
H_\textrm{orb,eq}^\textrm{Kerr} = \sqrt{ A^\textrm{eq}(r_c,a)\left(\mu^2 + \frac{p_r^2}{B^\textrm{eq}(r_c,a)} + \frac{p_\varphi^2}{r_c^2} \right)}.
\ee 
Here, $\mu$ denotes the mass of the test-particle\footnote{One of the features of the EOB formalism is that, after suitably deforming the Kerr Hamiltonian, it will be possible to replace $\mu$ by the reduced mass of the binary system, $\mu \equiv m_1m_2/(m_1+m_2)$, to describe the two-body effective Hamiltonian $H^\text{eff}$ entering Eq.~(\ref{eq:Heob}).}.
We see in Eq.~(\ref{eq:HorbKerr}) that the angular momentum dependence is encoded in the centrifugal term $p_\varphi^2/r_c^2$, involving the centrifugal radius $r_c$.
The construction of the EOB model of Ref.~\cite{Damour:2014sva} is based upon the idea of exploiting the role of $r_c$.
In addition, it was suggested to incorporate NLO spin-spin effects (though only for circular orbits) by redefining the relation between $r_c$ and the Boyer-Lindquist-like coordinate $r$, by adding to $a$ a new, radial dependent spin-quadratic term $\delta a^2(r)$.
This model can be extended without particular problems to equatorial, noncircular orbits. 
For example, the missing NLO spin-spin terms can be reproduced by a $p_r$-dependent term of the type 

\[
\left(1+ \frac{M \delta a^2_{p_r}}{r^3}\right)\frac{p_r^2}{B^\textrm{eq}}
\]
(where $\delta a^2_{p_r}$ is an appropriate quadratic combination of the individual spin parameters $a_1$ and $a_2$), or alternatively, by a modification of the $r$-$r_c$ relation inside of $B^\textrm{eq}$.

In the present work, our aim is to define an EOB dynamics which is able to give the simplest possible description of general, \emph{precessing} spinning binary systems with arbitrarily oriented spins.
When both spins, as well as the orbital plane, precess, there no longer exist useful analogs of the $z$-axis, and associated structures (equatorial plane, angular momentum $p_\varphi$) that motivated the emphasis on the centrifugal radius (\ref{eq:rcKerr}) and the associated form (\ref{eq:HorbKerr}) of the Kerr Hamiltonian.
This motivates us to reexamine the structure of the Kerr Hamiltonian when it is written in (Boyer-Lindquist-based) Cartesian-like coordinates $\bm{r} = (x,y,z)$, with $x= r\,\sin \theta\, \cos \varphi$, $y = r\, \sin \theta \, \sin \varphi$, $z = r\, \cos \theta$, namely:

\begin{widetext}
\begin{align}
\label{eq:KerrHam}
H_\textrm{orb}^\textrm{Kerr}  = & \sqrt{\frac{\Delta \lb r^2 + \na^2 \rb}{\mathcal{R}^4 + \Delta \na^2} \Bigg( \mu^2 + \frac{1}{1+ \frac{\na^2}{r^2}}\bigg[ \p^2 + \lb \frac{\Delta}{r^2} -1 \rb \np^2  - \frac{\lb r^2 + 2r + \na^2 \rb}{\mathcal{R}^4 + \Delta \na^2}  \enpa^2 \bigg] \Bigg)},
\end{align}
\end{widetext}

where $\bm{r} \equiv r\,\bm{n}$ and
\begin{align}
\Delta & =  r^2 - 2Mr + \av^2\\
\mathcal{R}^4 & = r^4 + r^2 \av^2 + 2Mr\av^2 = r^2r_c^2.
\end{align}
In this reformulation, the centrifugal term $p_\varphi^2/r_c^2$ has been split in two parts.
It is now contained in both the $\p^2$-contribution (with $\p^2 \equiv p_r^2 + p_\theta^2 / r^2 + p_\varphi^2/(r^2 \sin^2 \theta)$), and in the term $\enpa^2$ (which is equal to $\av^2 p_\varphi^2/r^2$ because $\av = |\av| \partial/\partial z$). 
Bringing these two parts together, and considering for simplicity equatorial orbits\footnote{Let us, however, recall in passing that $r_c$, Eq.~(\ref{eq:rcKerr}), continues to play a central role even for non equatorial orbits, modulo the introduction of a ``$\cos \theta$-dressing factor'', see Eq.~(2.2) in Ref.~\cite{Damour:2014sva}.} $\na = 0$, the centrifugal radius $r_c$ emerges from the identity 

\be
\frac{1}{r^2}\lb 1 - \frac{\av^2}{\mathcal{R}^4}\lb r^2 + 2r \rb \rb = \frac{1}{r_c^2}.
\ee
The Kerr Hamiltonian written as in Eq.~(\ref{eq:KerrHam}) will be the starting point of the new EOB model, i.e., we will look for an EOB effective, orbital Hamiltonian $H_\text{orb}^\text{eff}$ which is the simplest possible deformation of Eq.~(\ref{eq:KerrHam}).
Let us introduce specific notations for the coefficients of the various contributions as they appear in Eq.~(\ref{eq:KerrHam}), namely:
\begin{align}
	H_\textrm{orb}^\textrm{Kerr} & = \Big[ A^\text{Kerr} \Big( \mu^2 + B^\text{Kerr}_p \, \p^2 + B^\text{Kerr}_{np}\, \np^2 \nonumber \\
	& + B^\text{Kerr}_{\varepsilon n p}\, \enpa^2 \Big) \Big]^{1/2}.
	\label{eq:KerrHamSec}
\end{align}
We have thereby distinguished four principal sectors in $H^\text{Kerr}_\text{orb}$.
The first sector, described by the overall factor $A^\text{Kerr}(\bm{r},\av)$, is an anisotropic (spin-dependent) gravitational potential which generalizes the Schwarzschild (isotropic) potential $1 - 2M/r$.
It reads

	\begin{align}
	\label{eq:AKerr}
	 A^\text{Kerr}(\bm{r},\av)	& = \frac{\Delta \lb r^2 + \na^2 \rb}{\mathcal{R}^4 + \Delta \na^2}  \nonumber \\
					& = A^\text{Kerr, eq}(r_c)\frac{1+\frac{\na^2}{r^2}}{1+\frac{\Delta \na^2}{r^2 r_c^2}},
	\end{align}
where $A^\text{Kerr, eq}$ denotes the equatorial Kerr radial potential, given by

\be
\label{eq:AeqKerr}
A^\text{Kerr, eq}(r_c) = \lb 1 - \frac{2M}{r_c} \rb \frac{1+ \frac{2M}{r_c}}{1+ \frac{2M}{r}}.
\ee
As emphasized in \cite{Damour:2014sva}, $A^\text{Kerr, eq}(r_c)$ is a small deformation of $1- \frac{2M}{r_c}$, even for large spins.
The explicit expression of the remaining functions $B^\text{Kerr}_p$, $B^\text{Kerr}_{np}$ and $B^\text{Kerr}_{\varepsilon n p}$ can be deduced by a straightforward comparison with Eq.~(\ref{eq:KerrHam}), for instance $B_p^\text{Kerr} = 1/\lb 1+ \na^2/r^2\rb$.

We now take the square $\lb H_\textrm{orb}^\textrm{Kerr} \rb^2$  of the Kerr Hamiltonian, which is a quadratic function of the momenta, and investigate the momentum dependence of the spin-quadratic terms generated by each sector (without specifying the radial behavior $\sim 1/r^n$, $n \geq 3$).
More precisely, we formally expand the four separate building blocks $A^\text{Kerr}$, $B^\text{Kerr}_p$, $B^\text{Kerr}_{np}$ and $B^\text{Kerr}_{\varepsilon n p}$ in powers of $\av$ (keeping $\bm{r}$ fixed), and retain only the terms quadratic in spin (spin-spin terms).
We immediately observe that\begin{itemize}
	\item[i)] All momentum-independent terms $\av^2$ and $\na^2$ are encoded in the radial potential $A^\text{Kerr}(\mathbf{r},\mathbf{a})$.
	\item[ii)]The spin-spin terms contained in $B^\text{Kerr}_p\,\p^2$ and $B^\text{Kerr}_{np}\, \np^2$ can only be of the types $\p^2 \av^2$, $\p^2 \na^2$, and $\np^2 \av^2$, $\np^2 \na^2$, respectively. 
	\item[iii)] As the last contribution $B^\text{Kerr}_{\varepsilon n p} \, \enpa^2$ includes, as second factor, a term quadratic in $\av$, its spin-spin contribution only comes from the latter factor, namely $\enpa^2$.
	When decomposed in elementary scalar product factors, $\enpa^2$ is found to be a combination of six different terms: the four terms $\p^2 \av^2$, $\p^2 \na^2$, $\np \av^2$, $\np^2 \na^2$ that appeared in ii), together with two new couplings $\pa^2$ and $\np \na \pa$ (see Eq.~(3.9) of Ref.~\cite{Balmelli:2015lva}).
\end{itemize}
 
The fact that every sector plays a rather individual role suggests a natural procedure for including the NLO spin-spin coupling into a new EOB Hamiltonian.
This will be the topic of the next subsection.

\subsection{The Effective-One-Body orbital Hamiltonian}
\label{sec:Horbeff}

The idea at the basis of our new EOB Hamiltonian is to write the orbital part of the EOB effective Hamiltonian $H_\text{orb}^\text{eff}$ in the same form as Eq.~(\ref{eq:KerrHamSec}), but with (momentum-independent) coefficients $A(\bm{r},\nu, \bm{a}_1, \bm{a}_2)$, $B_p(\bm{r},\nu,\bm{a}_1, \bm{a}_2)$, $B_{np}(\bm{r},\nu,\bm{a}_1, \bm{a}_2)$ and $B_{\varepsilon n p}(\bm{r}, \nu,\bm{a}_1, \bm{a}_2)$ that are appropriate deformations of the coefficients $A^\text{Kerr}(\bm{r},\av)$, $B^\text{Kerr}_p(\bm{r},\av)$, $B^\text{Kerr}_{np}(\bm{r},\av)$ and $B^\text{Kerr}_{\varepsilon n p}(\bm{r},\av)$.
\newline
To be fully explicit, the structure of our new EOB Hamiltonian is given by Eq~(\ref{eq:Heob}), with $H^\text{eff}$ of the form Eq~(\ref{eq:orb+so}).
In the latter equation, the spin-orbit part is taken of the general form

\be
\label{eq:Hso}
H_\text{so} = G_S\,\bm{L}\cdot \bm{S} + G_{S^*}\, \bm{L}\cdot \bm{S^*},
\ee
in terms of the following symmetric combinations of the two spin vectors

\begin{align}
\label{eq:S}
\bm{S} \equiv & \,\bm{S}_1 + \bm{S}_2 \equiv m_1 \bm{a}_1 + m_2 \bm{a}_2, \\ 
\label{eq:S*}
\bm{S^*} \equiv &\, \mdu \bm{S}_1 + \mud \bm{S}_2 \equiv m_2 \bm{a}_1 + m_1 \bm{a}_2.
\end{align}
The factors $G_S$ and $G_{S^*}$ in Eq.~(\ref{eq:Hso}) are functions of $\bm{r}$, $\p$, $\av_1$ and $\av_2$, and are even in the spin vectors.
They are not the focus of the present work (see, however, below for more discussion of them).

In the present paper, we focus on a new definition of the spin-quadratic contribution of an effective orbital EOB Hamiltonian $H_\text{orb}^\text{eff}$ having the following structure:

\begin{align}
\label{eq:HorbeffStruct}
	H_\textrm{orb}^\textrm{eff} & = \Big[ A\lb \bm{r},\nu,\bm{a}_1, \bm{a}_2 \rb \Big( \mu^2 + B_p\lb \bm{r},\nu,\bm{a}_1, \bm{a}_2 \rb \p^2 \nonumber \\
	& +  B_{np}\lb \bm{r},\nu,\bm{a}_1, \bm{a}_2 \rb \np^2 \nonumber \\
	& +  B_{\varepsilon n p} \enpa^2\!\text{-like terms}+ Q_4 \Big) \Big]^{1/2},
\end{align}
where the structure of the last-indicated contribution on the right-hand-side (rhs) of Eq.~(\ref{eq:HorbeffStruct}) will be discussed below.

Let us start by specifying the structure that we shall require for the dependence of the EOB potentials $A$, $B_p$ and $B_{np}$ on the mass-ratio\footnote{We shall use here the convention $m_1 \geq m_2$ so that all the mass-ratios can be expressed in terms of $\nu = m_1m_2/(m_1+m_2)^2$.
E.g., $X_1 \equiv m_1/(m_1+m_2) = (1 + \sqrt{1-4\nu})/2$, $X_2 \equiv m_2/(m_1+m_2) = (1 - \sqrt{1-4\nu})/2$.}
$\nu$ and the two individual vectorial Kerr parameters of the two black holes $\bm{a}_1 \equiv \bm{S}_1/m_1$, $\bm{a}_2 \equiv \bm{S}_2/m_2$.
We recall \cite{Damour:2001tu} that an effective orbital Hamiltonian with the correct LO spin-spin coupling is simply obtained by replacing the Kerr spin vector $\av$ entering Eq.~(\ref{eq:KerrHam}) by the following effective spin vector

\be
\label{eq:effKerrParam}
\av_0 \equiv \bm{a}_1 + \bm{a}_2.
\ee
In addition to the replacement (\ref{eq:effKerrParam}), the two masses, $M$ and $\mu$, entering the Kerr dynamics are replaced by

\be
M = m_1 + m_2, \quad \quad \mu = \frac{m_2\,m_2}{m_2+m_2}.
\ee 
This suggests to look for EOB potentials $A$, $B_p$, $B_{np}$ of the form 

\begin{align}
\label{eq:deltaA}
A(\bm{r}, \nu, \bm{a}_1, \bm{a}_2) = \,& A^{\nu K_0}(\bm{r},\nu,\av_0) + \delta A, \\
\label{eq:deltaBp}
B_p(\bm{r}, \nu, \bm{a}_1, \bm{a}_2) = \,& B_p^{\nu K_0}(\bm{r},\nu,\av_0) + \delta B_p, \\
\label{eq:deltaBnp}
B_{np}(\bm{r}, \nu, \bm{a}_1, \bm{a}_2) = \,& B_{np}^{\nu K_0}(\bm{r},\nu,\av_0) + \delta B_{np},
\end{align}
where $A^{\nu K_0}$, $B_p^{\nu K_0}$, $B_{np}^{\nu K_0}$ are some $\nu$-deformed versions of the Kerr-like potentials defined by replacing $\av$ by $\av_0$ in the potentials $A^\text{Kerr}$, $B_p^\text{Kerr}$, $B_{np}^\text{Kerr}$ entering Eq.~(\ref{eq:KerrHamSec}), and where $\delta A$, $\delta B_p$, $\delta B_{np}$ are additional NLO spin-spin contributions.
Explicitly, we shall (following Ref.~\cite{Damour:2014sva}, except for the treatment of NLO spin-spin effects) take as $\nu$-deformed\footnote{For the purpose of this article, it is not necessary to be careful about the $\nu$-deformations of $A$ and $B_{np}$, since the NLO spin-spin coupling is not affected by them.
Indeed, neither $A$ nor $B_{np}$ contain $\nu$-dependent terms at the 1PN level, and thus there is no coupling of this type with the LO spin-spin part leading to NLO spin-spin terms.
However, an influence of the purely orbital $\nu$-deformation on the spin-spin sector is still present in the transformation between ADM and EOB coordinates, and also in the transformation between the effective and EOB Hamiltonians. 
}
, LO spin-spin, Kerr-like $A$ potential

\be
\label{eq:AnuK0}
A^{\nu K_0}(\bm{r}, \nu,\av_0) = A^\text{eq}(r_c,\nu,a_0) \frac{1 + \frac{\naz^2}{r^2}}{1+ \frac{\Delta(r,\av_0)\naz^2 }{r^2 r_c^2}},
\ee
where

\be
\label{eq:AeqBare}
A^\text{eq}(r_c,\nu,a_0) = A_\text{orb}(r_c,\nu) \frac{1 + \frac{2M}{r_c}}{1 + \frac{2M}{r}},
\ee
with

\be
\label{eq:Aorb}
A_\text{orb}(r_c,\nu) \equiv P^1_5\bigg[ A^\text{PN}_\text{orb} \lb\frac{M}{r_c}, \nu \rb \bigg],
\ee
where $P^1_5[A^\text{PN}_\text{orb}]$ denotes the $(1,5)$-Pad\'e resummation of the 5PN-level, Taylor-expanded orbital radial potential.
More precisely, we use Eqs.~(28)-(29) in \cite{Damour:2014sva} together with the exact value of $a_5^c(\nu)$ \cite{Bini:2013zaa} and the recent calibration $a_6^c(\nu)=3097.3 \nu^2 - 1330.6\nu + 81.38$ \cite{Nagar:2015xqa} (instead of the values for $a_5^c$ and $a_5^6$ that were employed in Ref.~\cite{Damour:2014sva}).

Here, and in the following, $r_c$ is defined as being the following function of $r$ and $\av_0$,

\be
\label{eq:rcEOB}
r_c \equiv \sqrt{r^2 + \av_0^2 + \frac{2M}{r}\av_0^2}.
\ee
As for the other Kerr-like EOB potentials, we take
\begin{align}
\label{eq:BpnuK0}
B_p^{\nu K_0} =\,& \frac{1}{1 +\frac{\naz^2}{r^2}}, \\
\label{eq:BnpnuK0}
B_{np}^{\nu K_0}  = \,&\frac{1}{1+\frac{\naz^2}{r^2}}\lb \frac{A^\textrm{eq}(r_c,\nu,a_0)}{D_\textrm{orb}(r_c,\nu)}\frac{r_c^2}{r^2}-1\rb,
\end{align}
where $A^\textrm{eq}(r_c,\nu,a_0)$ was defined in Eq.~(\ref{eq:AeqBare}) above, and where $D_\text{orb}(r_c,\nu)$ is defined by Eq.~(33) of \cite{Damour:2014sva} with $u_c \equiv M/r_c$.
Finally, the quartic-in-momenta term $Q_4$ that has to be added to the four main summands inside the effective Hamiltonian is defined by Eq.~(35) in Ref.~\cite{Damour:2014sva}.

\subsection{Canonical transformation from ADM to EOB}
\label{sec:Can}

In order to determine the additional, NLO spin-spin terms $\delta A$, $\delta B_p$, $\delta B_{np}$ in Eqs.~(\ref{eq:deltaA})-(\ref{eq:deltaBnp}) (as well as the NLO-accurate $B_{\varepsilon np} \enpa^2$-like terms in Eq.~(\ref{eq:HorbeffStruct})) we need to transform the ADM NLO spin-spin Hamiltonian $H_\text{ss}^\text{NLO(ADM)}$ \cite{Hergt:2008jn, Steinhoff:2008ji, Steinhoff:2007mb,Levi:2015msa} into a corresponding EOB Hamiltonian by means of a suitable canonical transformation.
As in Refs.~\cite{Balmelli:2013zna, Balmelli:2015lva}, this will be done by composing three successive canonical transformations.
The first transformation $G_\text{o}^\text{1PN}(\bm{r},\p)$ (given by Eqs.~(6.15)-(6.16) in Ref.~\cite{Buonanno:1998gg}) is of a purely orbital type, and has the following effect on spin-spin terms:
	\begin{equation}
	\label{eq:1sttransf}
		H_\text{ss}^{\text{NLO} \prime} = H_\text{ss}^\text{NLO(ADM)} + \big\{G_\text{o}^\text{1PN}, H_\text{ss}^\text{LO(ADM)} \big\}.	
	\end{equation}
It is followed by a LO spin-spin canonical transformation $G_\text{ss}^\text{LO}(\bm{r},\p, \bm{S}_1,\bm{S}_2)$ (given by Eq.~(5.15) in Ref.~\cite{Barausse:2009xi}, see also Eq.~(3.16) of Ref.~\cite{Balmelli:2015lva}) yielding  a further modification of spin-spin terms: 
	\begin{equation}
	\label{eq:2dtransf}
		H_\text{ss}^{\text{NLO} \prime \prime } = H_\text{ss}^{\text{NLO} \prime} + \big\{G_\text{ss}^\text{LO}, H_\text{o}^{\text{1PN}\prime} \big\},
	\end{equation}
	where 

	\be
	\label{eq:2dtransf2}
	H_\text{o}^{\text{1PN}\prime} = H_\text{o}^\text{1PN(ADM)} + \big\{G_\text{o}^\text{1PN}, H_\text{o}^\text{N(ADM)} \big\}.
	\ee
Finally, we perform a NLO spin-spin canonical transformation $G_\text{ss}^\text{NLO}(\bm{r},\p,\bm{S}_1,\bm{S}_2)$ (whose structure will be discussed below) yielding a last modification of spin-spin terms
	\begin{equation}
	\label{eq:3dtransf}
		H_\text{ss}^{\text{NLO} \prime \prime \prime} = H_\text{ss}^{\text{NLO} \prime \prime} + \big\{ G_\text{ss}^\text{NLO}, H_\text{N} \big\}.
	\end{equation}
$H_\text{ss}^{\text{NLO} \prime \prime \prime}$ must then be equal to the corresponding term in the PN expansion of the EOB Hamiltonian we are seeking.
It is convenient to focus the attention onto the squared effective orbital Hamiltonian $\lb H_\textrm{orb}^\textrm{eff} \rb^2$, which has an intuitive structure.
Because of the relation

\be
\label{eq:HEOBsq}
\hat{H}^\text{eff} = 1 + \hat{H}_\text{EOB}^\text{NR} + \frac{\nu}{2}\lb \hat{H}_\text{EOB}^\text{NR} \rb^2,
\ee
where $H_\text{EOB}^\text{NR} \equiv H_\text{EOB} - M$ is the ``non relativistic'' EOB Hamiltonian, and where the hat denotes a $\mu$-scaling $\hat{H} \equiv H/\mu$, $\hat{G} \equiv G/\mu$ we are left with the condition
\begin{align}
\label{eq:HeffSqcond}
	& \lb \hat{H}_\textrm{orb}^\textrm{eff} \rb^2 \Big|_\text{NLOss} =  \nonumber \\
	&  2 \lb \hat{H}_\text{ss}^{\text{NLO} \prime \prime \prime} + \lb 1 + \nu \rb \hat{H}_\text{N} \lb \hat{H}_\text{ss}^\text{LO(ADM)} + \{ \hat{G}_\text{ss}^\text{LO}, \hat{H}_\text{N} \} \rb \rb,
\end{align}
where the notation on the left hand side simply denotes the NLO spin-spin part of the PN expansion of $ \lb \hat{H}_\textrm{orb}^\textrm{eff}\rb ^2$.
In other words, our problem is to find a suitable $G_\text{ss}^\text{NLO}$ such that the rhs of Eq.~(\ref{eq:HeffSqcond}) is equal to the NLO spin-spin contribution to the expression

\begin{align}
	\lb H_\textrm{orb}^\textrm{eff} \rb^2 & = \Big[ \lb A^{\nu K_0} + \delta A \rb\Big( \mu^2 + \lb B_p^{\nu K_0} + \delta B_p \rb \p^2 \nonumber \\
	& +  \lb B_{np}^{\nu K_0} + \delta B_{np} \rb \np^2 \nonumber \\
	& +  B_{\varepsilon n p} \enpa^2\!\text{-like terms}+ Q_4 \Big) \Big]^{1/2},
\end{align}
with appropriate NLO spin-spin terms $\delta A$, $\delta B_p$, $\delta B_{np}$, and with a suitable NLO-accurate EOB version of the $\enpa^2$ term in the Kerr Hamiltonian (\ref{eq:KerrHam}).

We introduce at this point a change in the notation.
Since NLO spin-spin terms are more conveniently expressed by dimensionless quantities, we will from now on only make use of the dimensionless rescaled variables $\hat{r} \equiv r/M$, $\hat{r}_c \equiv r_c/M$, $\hat{\p} \equiv \p/\mu$, $\cu \equiv \bm{a}_1/m_1$, $\cd \equiv \bm{a}_2 /m_2$, $\cz \equiv \az/M$, $\hat{H} \equiv H/\mu$ and $\hat{G} \equiv G/\mu$.
However, in order to lighten the notation, we will omit to display the hats on the dynamical variables $r$, $r_c$ and $\p$. 

Before evaluating Eq.~(\ref{eq:HeffSqcond}), it is necessary to specify the form of the canonical transformation (\ref{eq:3dtransf}).
In Ref.~\cite{Balmelli:2015lva}, the generating function $\hat{G}_\text{ss}^\text{NLO}$ had been chosen in a rather general way, which involved terms cubic in the momenta.
The latter terms gave rise, in the Hamiltonian, to NLO spin-spin terms that were quartic in the momenta.
The presence of such terms is a feature not shared by the ADM Hamiltonian, but was related to the idea of defining, in the EOB formalism, an ``effective spin'' that may also depend on $\p^2$ and $\np^2$, thereby introducing higher powers of the momenta. 

In this paper, by contrast, we want to hold the dependence on the momenta as simple as possible.
We found it possible to end up with a squared effective EOB Hamiltonian involving only quadratic-in-momenta spin-spin terms by choosing an NLO spin-spin generating function $\hat{G}_\text{ss}^\text{NLO}$ which is only linear in momenta (rather than cubic as in Ref.~\cite{Balmelli:2015lva}).
[This fact relies on the combined structure of the LO spin-spin canonical transformation $G_\text{ss}^\text{LO}$ \cite{Barausse:2009xi} (going from ADM coordinates to Boyer-Lindquist coordinates) and of the nonlinear transformation relating the effective Hamiltonian to the real one.]
Among the 33 gauge coefficients taken into account in Ref.~\cite{Balmelli:2015lva} for $\hat{G}_\text{ss}^\text{NLO}$, we only need to maintain 10 of them.\footnote{The 23 coefficients that we discard here are all those cubic in $\p$.
Each of them leads, after the Poisson Bracket with the Newtonian Hamiltonian, to terms quartic in the momenta. 
An explicit calculation easily shows that the so obtained 23 quartic expressions are linearly independent in the 32-dimensional space of NLO spin-spin polynomials that are quartic in the momenta, whose basis is defined by scalars of the type $\p^4 \cicj/r^2$, $\np^4 \cicj/r^2$, and so on.
There is therefore no way of tuning these 23 coefficients, apart from setting all of them to zero, that prevents the transformed Hamiltonian from being quartic in the momenta.}
We thus consider a generating function of the following form:\footnote{We warn the reader that the nomenclature of the gauge coefficients differs significantly from the one used in Refs.~\cite{Balmelli:2013zna,Balmelli:2015lva}.
In particular, the coefficients $\alpha$, $\beta$ and $\gamma$ used here correspond to $\gamma^{(\chi)}$, $\gamma^{(n)}$ and $\gamma^{(np)}$ in Ref.~\cite{Balmelli:2015lva}.
The reason beyond these choices has been that of favoring the readability and self-consistence of this paper over the continuity with respect to Ref.~\cite{Balmelli:2015lva}.}

\be
\label{eq:GssNLO}
\begin{split}
\hat{G}_\text{ss}^\text{NLO} = 	& \frac{\np}{r^2} \Big( \alpha_{ij} \cicj + \beta_{ij} \nci \ncj \Big)\\ 
							&+ \frac{1}{r^2}\gamma_{ij} \nci \pcj ,
\end{split}
\ee
where we use the summation convention on the spin labels $i,j=1,2$, and where the coefficients $\alpha_{ij}$ and $\beta_{ij}$ are assumed to be symmetric, while $\gamma_{ij} \neq \gamma_{ji}$.

The change induced by $\hat{G}_\text{ss}^\text{NLO}$ in the Hamiltonian is

\begin{widetext}
\be
\label{eq:deltaH}
\begin{split}
	  \big\{ \hat{G}_\text{ss}^\text{NLO}, \hat{H}_\text{N} \big\}  =  &\frac{1}{r^3} \Big[ \lb \alpha_{ij} \p^2 - 3\alpha_{ij}\np^2 - \frac{\alpha_{ij}}{r}  \rb \cicj  + \lb \beta_{ij} \p^2 - 5\beta_{ij}\np^2 - \frac{\beta_{ij}  + \gamma_{(ij)}}{r}  \rb \nci \ncj \\
		& + \gamma_{(ij)} \pci \pcj + \lb2\beta_{ij} -3\gamma_{ij}\rb \np \nci \pcj \Big],
\end{split}
\ee
\end{widetext}
where we have introduced the symmetrized coefficients $\gamma_{(ij)} \equiv(\gamma_{ij} + \gamma_{ji})/2$ in order to point out that the only term which is not symmetric under exchange of the indices $i$ and $j$ is the last one, i.e., $-3\gamma_{ij} r^{-3} \np \nci \pcj$.
We will show in the next subsection why $\gamma_{ij}$ must contain an antisymmetric part $\gamma_{[ij]}$, and how $\gamma_{[ij]}$ can be used to yield a simple $H_\text{orb}^\text{eff}$.

\subsection{Gauge choice}
\label{sec:Gauge}

One of the useful features of the EOB formalism is to use canonical transformations as gauge transformations able (after some gauge choice) to simplify the structure of PN-expanded Hamiltonians.
Here, we shall apply this philosophy to the NLO spin-spin Hamiltonian.
The original NLO spin-spin Hamiltonian, obtained in ADM gauge in Refs.~\cite{Hergt:2008jn, Steinhoff:2008ji, Steinhoff:2007mb}, contains 25 different terms in the center-of-mass frame (see Eq.~(2.9a) of Ref.~\cite{Balmelli:2013zna}, which accounts for both spin(1)-spin(1) and spin(2)-spin(2) terms, and Eq.~(3.15) of Ref.~\cite{Balmelli:2015lva} (spin(1)-spin(2)) for a center-of-mass formulation). 
[This is the generic number of terms for an NLO spin-spin Hamiltonian which is at most quadratic in momenta, as the ADM spin-spin Hamiltonian happens to be.]
As we have introduced in Eq.~(\ref{eq:GssNLO}) a NLO spin-spin transformation involving 10 arbitrary parameters ($\alpha_{(ij)}$, $\beta_{(ij)}$, $\gamma_{(ij)}$ and $\gamma_{[12]}$), we expect to be able to end up with a simplified EOB NLO spin-spin Hamiltonian containing at most 15 different terms.
In particular, we wish to simplify the \emph{a priori} most complicated sector of the ADM Hamiltonian (and of its generic EOB counterpart), namely the sector comprising the seven different terms

\be
\label{eq:SevenTerms}
\pci \pcj \quad \text{and} \quad \np \nci \pcj
\ee
appearing in the last two contributions on the rhs of Eq.~(\ref{eq:deltaH}).
As discussed above, in the Kerr case (with only one $\bm{\chi}$), these couplings came out of the decomposition of the Kerr coupling $B_{\varepsilon np} \enpa^2$ into elementary product factors.
We found convenient to use the freedom of $\hat{G}_\text{ss}^\text{NLO}$ to impose that the EOB sector containing the seven different terms (\ref{eq:SevenTerms}) take the following maximally simplified form:

\be
\label{eq:BenpKerr}
B_{\varepsilon np}^\text{Kerr} (\bm{r}, \av_0) \lb ( \bm{n} \times \p) \cdot \av_0 \rb^2
\ee
differing by its Kerr counterpart (last terms on the rhs of Eq.~(\ref{eq:KerrHam})) only by the replacement $\av \to \av_0 \equiv \av_1 + \av_2$.
It is easily checked that this requirement uniquely fixes 7 degrees of freedom in $\hat{G}_\text{ss}^\text{NLO}$, in determining the gauge parameters $\beta_{(ij)}$ and $\gamma_{ij}$ (which, as exhibited in Eq.~(\ref{eq:deltaH}), entered the gauge variation of the seven terms (\ref{eq:SevenTerms})).

More precisely, these 7 gauge parameters must take the values

\begin{subequations}
\begin{align}
	\beta_{11}  	& = - \lb \frac{1}{2} + \frac{3}{4}\nu \rb \lb X_1 -\nu \rb  \\
	\beta_{22} 		& = - \lb \frac{1}{2} + \frac{3}{4}\nu \rb \lb X_2 -\nu \rb  \\
	\beta_{12}		& =\, \beta_{21} = - \lb \frac{1}{2} + \frac{3}{4}\nu \rb \nu
\end{align}
\end{subequations}
and
\begin{subequations}
\begin{align}
	\gamma_{11}  	& = X_1 - \nu - \frac{\nu^2}{4} \\
	\gamma_{22} 	& = X_2 - \nu - \frac{\nu^2}{4} \\
	\gamma_{12} 	& = \frac{\nu}{2} X_1 - \frac{\nu^2}{4} \\
	\gamma_{21} 	& = \frac{\nu}{2} X_2 - \frac{\nu^2}{4}.
\end{align} 
\end{subequations}
Note that, in the limit $m_2 \ll m_1$ (under which $X_2 \to 0$, $X_1 \to 1$, $\nu \to 0$) we have $\beta_{11} \to - \frac{1}{2}$ and $\gamma_{11} \to 1$, which is a necessary requirement for the structure of $\hat{G}_\text{ss}^\text{NLO}$ (as discussed in Refs.~\cite{Balmelli:2013zna, Balmelli:2015lva}).
Note also that the antisymmetric part of $\gamma_{ij}$ is fixed to the value

\be
\label{eq:gamma_antis}
\gamma_{[ij]} = \frac{\nu}{4}\lb X_i - X_j\rb.
\ee

It is easily checked (using Eq.~(\ref{eq:deltaH})) that this value allows one to gauge away the antisymmetric-looking\footnote{Note, however, that this term is symmetric under the combined permutation $X_1 \leftrightarrow X_2$, $\cu \leftrightarrow \cd$.} ADM term \cite{Steinhoff:2007mb} 

	\be
	\begin{split}
		\hat{H}_\text{ss, antis.}^\text{NLO(ADM)} =\,& \frac{3}{4}\nu \frac{\np}{r^3} (X_1-X_2)\big( \ncu \pcd \\
				& - \ncd \pcu \big),
	\end{split}
	\ee
so as to end up with a symmetric contribution $\propto \ncu \pcd + \ncd \pcu$ of the type contained in the expansion of the term $\enpaz^2$.

Having fixed the $B_{\varepsilon np} \enpa^2$ sector by using the 7 gauge parameters $\beta_{(ij)}$ $\gamma_{ij}$, we are left with the 3 gauge parameters $\alpha_{(ij)}$ to simplify the NLO contributions $\delta A$, $\delta B_p$ and $\delta B_{np}$ to the remaining physical sectors of the NLO spin-spin EOB Hamiltonian.
As we started from 25 different contributions and used only 7 gauge parameters, we would expect $\delta A$, $\delta B_p$ and $\delta B_{np}$ to involve $25-7=18$ different contributions, in the form of 6 different quadratic forms in the two spin vectors.
More specifically, one can \emph{a priori} decompose $\delta A$, $\delta B_p$ and $\delta B_{np}$ in the form

\begin{align}
&	\delta A 		=  \frac{1}{r^4}\lb A^Q_{\chi} - A^Q_{n \chi}\rb \\
&	\delta B_{p}	=  \frac{1}{r^3}\lb B^Q_{p,\chi} - B^Q_{p,n \chi}\rb \\
&	\delta B_{np} 	=  \frac{1}{r^3}\lb B^Q_{np,\chi} - B^Q_{np,n \chi}\rb,
\end{align}
(where the minus signs are introduced for later convenience) with six (symmetric) quadratic forms

\begin{align}
&	A^Q_{\chi}		  = a^{\chi}_{ij} \cicj \\
&	A^Q_{n \chi}		  = a^{n \chi}_{ij} \nci \ncj \\ 
&	B^Q_{p, \chi}	  = b^{p,\chi}_{ij} \cicj \\
&	B^Q_{p, n\chi}	  = b^{p, n\chi}_{ij} \nci \ncj \\ 
&	B^Q_{np, \chi}	  = b^{np,\chi}_{ij} \cicj \\
&	B^Q_{np, n\chi}	  = b^{np, n\chi}_{ij} \nci \ncj. 
\end{align}
[Note that the summation convention on the indices $i$,$j$ means that, e.g., $A^Q_\chi =  a_{11}^\chi \cu^2 + 2a_{12}^\chi \cucd + a_{22}^\chi \cd^2$.]
A first remarkable finding is that our request of having the simple, Kerr-like form (\ref{eq:BenpKerr}) implies another simplification for free.
Namely, we find that the 3 coefficients 

\be
\label{eq:bnpnc}
b_{ij}^{np,n\chi} = 0,
\ee
so that the second quadratic form, $B^Q_{np,n\chi}$, entering $\delta B_{np}$ simply vanishes.
We also find that the coefficients of the second quadratic forms  $A^Q_{n\chi}$ and $B^Q_{p, n\chi}$ entering $\delta A$ and $\delta B_p$  are uniquely fixed to the values

\begin{subequations}
\begin{align}
	a_{11}^{n\chi}	= &  \lb 2\nu X_1 + \frac{5}{2}\nu^2 \rb   \\
	a_{22}^{n\chi}	= &  \lb 2\nu X_2 + \frac{5}{2}\nu^2 \rb   \\
	a_{12}^{n\chi}	= &\, a_{21}^{n\chi} =   \lb \frac{3}{2}\nu - \frac{7}{2}\nu^2\rb 
\end{align}
\end{subequations}
\begin{subequations}
\label{eq:bpnc}
\begin{align}
	b_{11}^{p, n\chi} = &  \lb 9 \nu X_1 -\frac{15}{4} \nu^2 \rb   \\
	b_{22}^{p, n\chi} = &  \lb 9 \nu X_2 - \frac{15}{4} \nu^2 \rb  \\
	b_{12}^{p, n\chi} = &\, b_{21}^{p, n\chi} =   \lb 3\nu + \frac{9}{4} \nu^2  \rb.
\end{align} 
\end{subequations}
Let us now consider the three remaining quadratic forms (linear in $\cicj$) $A^Q_\chi$, $B^Q_{p,\chi}$ and $B^Q_{np,\chi}$.
These three forms are not fixed by our previous request, because they depend on the three gauge parameters $\alpha_{(ij)}$, which are still free at this stage.
In view of Eq.~(\ref{eq:deltaH}) (keeping in mind the factor 2 in Eq.~(\ref{eq:HeffSqcond})) the effect of a gauge shift $\delta \alpha_{ij}$ on the three quadratic forms $A^Q_\chi$, $B^Q_{p,\chi}$ and $B^Q_{np,\chi}$ is

\begin{align}
&	\delta A^Q_{\chi} 		 = -2\, \delta \alpha_{ij} \cicj \\ 
&	\delta B^Q_{p, \chi}		 = 2\, \delta \alpha_{ij} \cicj \\ 
&	\delta B^Q_{np, \chi}	 = -6\, \delta \alpha_{ij} \cicj. 
\end{align}
In view of these transformation properties we could use the $\alpha_{ij}$-freedom to set to zero any of the three forms $A^Q_\chi$, $B^Q_{p,\chi}$ and $B^Q_{np,\chi}$.
Setting to zero $A^Q_\chi$ does not seem physically appealing because $A^Q_\chi$ has a relatively simple and intuitive meaning as a higher-order contribution to the already present spin-spin contribution to the radial potential $A^{\nu K_0}$, Eq.~(\ref{eq:AnuK0}).
This leaves us with two natural options: setting either $B^Q_{p,\chi}$ or $B^Q_{np,\chi}$ to zero.

Let us first briefly discuss the latter option, i.e. using $\alpha_{ij}$ to set $b_{ij}^{np,\chi} \equiv 0$.
Explicit calculations then show that a simple link emerges between the resulting gauge-fixed $B^Q_{p,\chi}$ and the form $B^Q_{p,n\chi}$ which was already fixed (and given by Eq.~(\ref{eq:bpnc})).
Indeed, we find in this case that the following relation holds

\be
\label{eq:bpc}
b_{ij}^{p,\chi} = \frac{1}{3}b_{ij}^{p, n \chi}.
\ee
This relation means that the momentum-dependent part of the NLO spin-spin contribution to $\lb H^\text{eff}\rb^2$ takes the simple form

\[
\frac{\p^2}{r^3}b_{ij}^{p,\chi}\lb \cicj - 3\nci \ncj \rb, 
\]
where we recognize a coupling between $\p^2$ and a spin-spin structure akin to the LO quadrupole potential present in the ADM Hamiltonian

\be
\hat{H}_\text{ss}^\text{LO(ADM)} = -\frac{1}{2r^3}\lb\cz^2 - 3 \ncz^2 \rb  = \frac{\cz^2}{r^3} P_2(\cos \vartheta).
\ee
In the last equality, $\vartheta$ is the angle between $\bm{n}$ and $\cz$, and $P_2$ is the second Legendre polynomial.
Notice that a coupling of the type $\hat{H}_\text{N} \hat{H}_\text{ss}^\text{LO(ADM)}$ (which involves $\p^2 \hat{H}_\text{ss}^\text{LO(ADM)}$) is explicitly visible in Eq.~(\ref{eq:HeffSqcond}).

The other option is to use the $\alpha_{ij}$ freedom to set, instead, the form $B^Q_{p, \chi}$ to zero, i.e.

\be
\label{eq:DJS}
b_{ij}^{p,\chi} \equiv 0.
\ee
In analogy to Refs.~\cite{Damour:2008qf,Nagar:2011fx}, this choice can be called a Damour-Jaranowski-Sch\"afer gauge.
When the orbits are circular and equatorial, the gauge-choice (\ref{eq:DJS}) leads to a very simple spin-spin structure, since in that case  $A^Q_{\chi}$ becomes the only quadratic form that does not vanish.
Consequently, all new NLO spin-spin information is contained in the radial potential $A$.
We will adopt this gauge for the rest of the paper. 

To satisfy Eq.~(\ref{eq:DJS}), the $\alpha_{ij}$ gauge parameters must be taken to be

\begin{subequations}
\begin{align}
	\alpha_{11} 	= 	& \; -\lb \frac{1}{2} + \frac{5}{4}\nu \rb X_1 + \frac{\nu}{2} + \frac{\nu^2}{2}    \\
	\alpha_{22}	=	& \; -\lb \frac{1}{2} + \frac{5}{4}\nu \rb X_2 + \frac{\nu}{2} + \frac{\nu^2}{2}    \\
	\alpha_{12}	=	& \; \alpha_{21} = -\frac{\nu}{2}.
\end{align}
\end{subequations}
In the limit $m_2 \ll m_1$, we have $\alpha_{11} \to -\frac{1}{2}$, which is a necessary requirement for the structure of $\hat{G}_\text{ss}^\text{NLO}$ \cite{Balmelli:2013zna, Balmelli:2015lva}.
Solving Eq.~(\ref{eq:HeffSqcond}) then leads first to

\begin{subequations}
\begin{align}
	a_{11}^{\chi} 	= 	& \; 3\nu X_1 - \frac{\nu^2}{2}   \\
	a_{22}^{\chi}	=	& \; 3\nu X_2 - \frac{\nu^2}{2}   \\
	a_{12}^{\chi}	=	& \; a_{21}^{\chi} = \nu - \frac{\nu^2}{2}
\end{align}
\end{subequations}
and then to a remarkable result for the coefficients of $B^Q_{np,\chi}$.
Namely, we find that they turn out to coincide with the coefficients of the above-determined quadratic form $B^Q_{p,n\chi}$, i.e.

\be
\label{eq:bnpc}
	b_{ij}^{np, \chi} =  b_{ij}^{p, n \chi}.
\ee
Here, as in the case of the other possible gauge $b_{ij}^{np,\chi} \equiv 0$, a symmetry becomes visible between $b_{ij}$-type coefficients belonging to different quadratic forms.

The final result is remarkable: the information stored in the 9 coefficients $a_{ij}^{\chi}$, $a_{ij}^{n \chi}$ and $b_{ij}^{p, n \chi}$ is sufficient, once inserted in the EOB Hamiltonian, to reproduce the whole NLO spin-spin coupling (which initially involved 25 different terms).
The EOB has not only exploited the full power of the gauge transformations, involving 10 parameters, but has also revealed 6 additional and unexpected symmetries (see the Appendix for a further discussion of these symmetries). 
Notice that the EOB Hamiltonian proposed in Ref.~\cite{Balmelli:2015lva} involved 12 different terms.
A symmetry similar to (\ref{eq:bnpnc}) was present, but there was no equivalent to (\ref{eq:bpc}) or (\ref{eq:bnpc}).

To summarize the results so far, the effective orbital Hamiltonian has the form
\begin{widetext}
\begin{align}
\label{eq:Horbeff}
\hat{H}_\textrm{orb}^\textrm{eff}  = & \sqrt{A \Bigg(1 + B_p\, \p^2 + B_{np}\, \np^2  - \frac{1}{1 + \frac{\ncz^2}{r^2}}\frac{\lb r^2 + 2r + \ncz^2 \rb}{\mathcal{R}^4 + \Delta \ncz^2}  \enpcz^2  + Q_4 \Bigg)}.
\end{align}

\end{widetext}

Here, the quantities entering the $\enpaz^2$ term are

\begin{align}
\Delta 			& =  r^2 - 2r + \cz^2\\
\mathcal{R}^4 	& = r^4 + r^2 \cz^2 + 2r\,\cz^2,
\end{align}
with the dimensionless effective spin

\be
\cz 			= X_1 \cu + X_2 \cd. 
\ee
On the other hand, we obtained above explicit, but non-resummed, expressions for the NLO-spin-spin accurate potentials $A$, $B_p$ and $B_{np}$.
In our preferred ($B^Q_{p,\chi}=0$) gauge, and in view of the remarkable cancellation of $B^Q_{np,n\chi}$, they have the form

\begin{align}
A(\bm{r},\nu,\cu,\cd) =\,& A^{\nu K_0} + \frac{1}{r^4}\lb A^Q_\chi - A^Q_{n\chi} \rb,\\
B_p(\bm{r},\nu,\cu,\cd) =\,& B_p^{\nu K_0} - \frac{1}{r^3}B^Q_{n\chi},\\
B_{np}(\bm{r},\nu,\cu,\cd) =\,& B_{np}^{\nu K_0} + \frac{1}{r^3}B^Q_{\chi}.
\end{align}
Here, $A^{\nu K_0}$, $B_p^{\nu K_0}$, $B_{np}^{\nu K_0}$ have been defined in Eqs.~(\ref{eq:AnuK0}), (\ref{eq:BpnuK0}), (\ref{eq:BnpnuK0}), while the four remaining NLO spin-spin quadratic forms entering our results (here and henceforth we simplify the notation by suppressing the index $p$ on $B^Q_{p,n\chi}$ and the index $np$ on $B^Q_{np,\chi}$) take the following explicit form:

\begin{widetext}
\begin{align}
\label{eq:A^Qc}
& A^Q_{\chi} 	= \lb 3\nu X_1 - \frac{\nu^2}{2}\rb \cu^2 + \lb 3\nu X_2 - \frac{\nu^2}{2}\rb \cd^2 + \lb 2\nu - \nu^2\rb \cucd \\
\label{eq:A^Qnc}
& A^Q_{n \chi} 	= \lb 2\nu X_1 + \frac{5}{2}\nu^2 \rb \ncu^2 + \lb 2\nu X_2 + \frac{5}{2}\nu^2 \rb \ncd^2 + \lb 3 \nu - 7 \nu^2\rb \ncu \ncd \\
\label{eq:B^Q}
& B^Q_{\chi} 	= \lb 9 \nu X_1 -\frac{15}{4} \nu^2 \rb \cu^2  + \lb 9 \nu X_2 - \frac{15}{4} \nu^2 \rb \cd^2 + \lb 6\nu + \frac{9}{2} \nu^2  \rb \cucd \\
\label{eq:B^Qnc}
& B^Q_{n\chi} 	= \lb 9 \nu X_1 -\frac{15}{4} \nu^2 \rb \ncu^2  + \lb 9 \nu X_2 - \frac{15}{4} \nu^2 \rb \ncd^2 + \lb 6\nu + \frac{9}{2} \nu^2  \rb \ncu \ncd. 
\end{align}  

\end{widetext}
Note again the remarkable fact, found above, Eq.~(\ref{eq:bnpc}), that the coefficients of $B^Q_{n\chi}$ coincide with the coefficients of $B^Q_\chi$ (i.e. $B^Q_{n \chi}$ is obtained from $B^Q_{\chi}$ simply by replacing $\cicj \to \nci \ncj$).

\subsection{Resummation options}
\label{sec:resum}

We wish to discuss now various options for incorporating the NLO spin-spin contributions $r^{-4} \lb A^Q_\chi - A^Q_{n\chi} \rb$, $-r^{-3}B^Q_{n\chi}$ and $r^{-3}B^Q_{\chi}$ in a somewhat resummed manner, within the $\nu$-deformed Kerr-like basic contributions $A^{\nu K_0}$, $B_p^{\nu K_0}$ and $B_{np}^{\nu K_0}$.
Let us first consider the contributions $\propto A^Q_{n\chi}$ and $B^Q_{n\chi}$, which are quadratic in $\nci$.
The presence in $A^{\nu K_0}$, Eq.~(\ref{eq:AnuK0}), of a factor $1 + \ncz^2/r^2$ and in $B_p^{\nu K_0}$, Eq.~(\ref{eq:BpnuK0}), of a factor $\lb 1 + \ncz^2/r^2\rb^{-1}$ suggests to incorporate the quadratic forms $ r^{-4}A^Q_{n\chi}$ and $r^{-3}B^Q_{n\chi}$ as additive modifications of the term $r^{-2} \ncz^2$.
This leads to the forms

\be
\label{eq:Anew}
A(\bm{r},\nu,\cu,\cd) \equiv A^\textrm{eq}(r_c,\nu,\cicj)\frac{1+\frac{\ncz^2}{r^2} - \frac{A^Q_{n \chi}}{r^4}}{1+\frac{\Delta \ncz^2}{r^2 r_c^2}},
\ee
and 
\be
\label{eq:Bpnew}
B_p(\bm{r},\nu,\cu,\cd) \equiv  \frac{1}{1 + \frac{\ncz^2}{r^2} + \frac{B^Q_{n \chi}}{r^3}}.
\ee
We recall that, in this work, the centrifugal radius is defined as

\be
\label{eq:rccz}
r_c = \sqrt{r^2 + \cz^2 + \frac{2 \cz^2}{r}}.
\ee
In Eq.~(\ref{eq:Anew}) we have introduced the notation  $A^\textrm{eq}(r_c,\nu,\cicj)$ for an equatorial potential (remaining in the limit $\nci \to 0$) which should incorporate, in a combined manner, both the Kerr-like equatorial potential (\ref{eq:AeqBare}) and the purely radial NLO spin-spin correction $r^{-4}A^Q_\chi$.
There are two main possibilities for doing so:

\begin{itemize}
	\item[i)] A full factorization 
	\be
	\label{eq:AeqnewFact}
	A^\text{eq}(r_c,\nu,\cicj) \equiv A_\text{orb}(r_c,\nu) \frac{1 + \frac{2}{r_c}}{1 + \frac{2}{r}}\lb 1 + \frac{A^Q_{\chi}}{r_c^4} \rb.
	\ee	

	\item[ii)] A semi-additive inclusion 
	\be
	\label{eq:Aeqnew}
	A^\text{eq}(r_c,\nu,\cicj) \equiv A_\text{orb}(r_c,\nu) \frac{1 + \frac{2}{r_c}+ \frac{A^Q_{\chi}}{r_c^4}}{1 + \frac{2}{r}}.
	\ee	
\end{itemize}
Here, $A_\text{orb}(r_c,\nu)$ denotes the Pad\'e-resummed orbital potential (\ref{eq:Aorb}), which entered the Kerr-like equatorial potential (\ref{eq:AeqBare}).
Note that the option ii) is equivalent to replacing the factor $1+A^Q_\chi/r_c^4$ of option i) by $1+A^Q_\chi/(r_c^4 + 2r_c^3)$.
As a consequence, the second option reduces the effect of $A^Q_\chi$ compared to the first option.
In addition, let us recall that the factor $(1+2/r_c)/(1+2/r)$ in $A^\text{eq}(r_c)$ is smaller than 1 and embodies the attractive nature of the extra coupling linked to the combined effect of the quadrupole deformations and of the spin(1)-spin(2) interaction
\be
\frac{1 + \frac{2}{r_c}}{1+ \frac{2}{r}} \approx 1 - \frac{\cz^2}{r_c^3} + ...
\ee
We then see that the main effect, for equatorial orbits, of NLO spin-spin effects is to reduce the attractive character of the LO spin-spin coupling by adding a \emph{repulsive} coupling $\propto + A^Q_\chi/r^4$.
[We will see in the next subsection that, in most cases, $A^Q_\chi$ is positive.]

Alternative versions ib) and iib) of the above options can be obtained by using the Boyer-Lindquist radius instead of the centrifugal one, thus substituting $A^Q_{\chi}/r_c^4$ with $A^Q_{\chi}/r^4$.
Among these four options, we choose in the following the semi-additive inclusion ii), given by Eq.~(\ref{eq:Aeqnew}), as our standard one.

Let us finally consider various ways of incorporating the correction $r^{-3} B^Q_\chi$ in the Kerr-like basic potential $B_{np}^{\nu K_0}$, Eq.~(\ref{eq:BnpnuK0}).
A simple way is to modify the fraction $r_c^2/r^2$ as it appears in Eq.~(\ref{eq:BnpnuK0}).
We choose here to do it by defining
\be
\label{eq:Bnpnew}
B_{np}  \equiv \frac{1}{1+\frac{\ncz^2}{r^2}}\lb \frac{A_B^\textrm{eq}(r_c)}{D_\textrm{orb}}\frac{r_c^2 + \frac{B^Q_{\chi}}{r}}{r^2} -1\rb,
\ee
where we used a ``bare'' version $A_B^\textrm{eq}(r_c)$ of the equatorial radial potential (i.e., a version which does not contain the insertion of $A^Q_\chi$), namely  

\be
\label{eq:A_B}
A_B^\text{eq}(r_c,\nu,a_0) \equiv A_\text{orb}(r_c,\nu)\frac{1 + \frac{2M}{r_c}}{1+ \frac{2M}{r}}.
\ee

\subsection{The quadratic forms}
\label{sec:quad}

\begin{figure}
	\centering
	\includegraphics{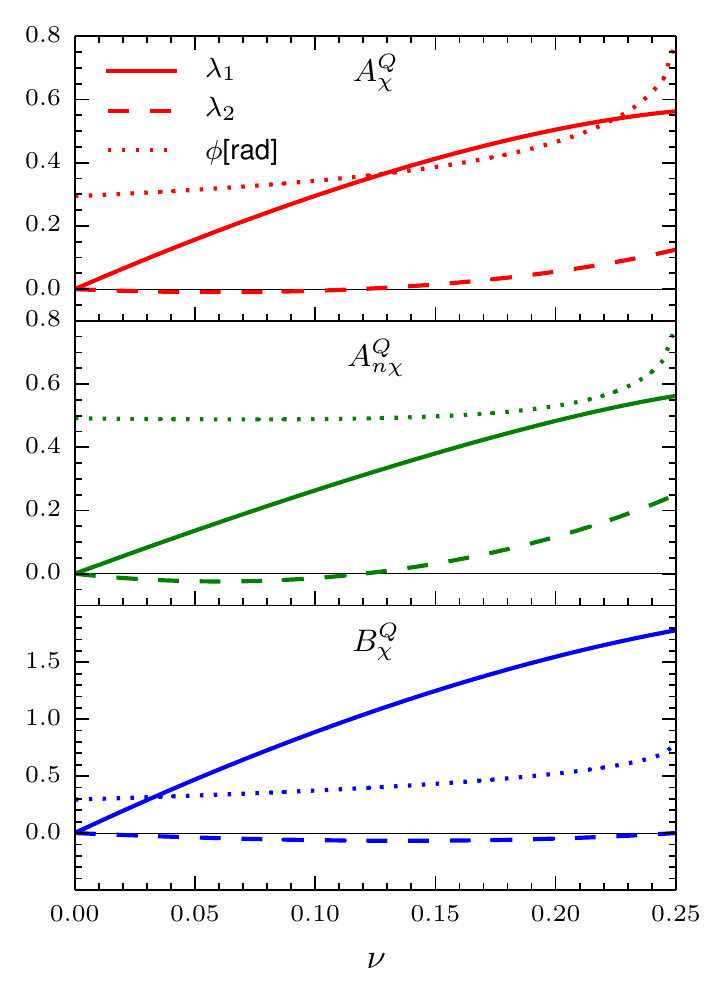}
	\caption{The eigenvalues $\lambda_1$, $\lambda_2$ and the rotation angle $\phi$ are plotted as a function of $\nu$ for the quadratic forms $A^Q_{\chi}$, $A^Q_{n \chi}$ and $B^Q_{\chi}$. 
The information relative to the form $B^Q_{n \chi}$ is equivalent to the one provided by the plot of $B^Q_{\chi}$.
Notice that $\phi(1/4) = \pi/4 \approx 0.79$ for all forms. }
	\label{fig:QuadEig}
\end{figure}

To have a feeling for the physical effects of the various NLO spin-spin quadratic forms $A^Q_\chi$, $A^Q_{n\chi}$, $B^Q_\chi$ entering our results, we investigate here their magnitudes and their signs as functions of the two spins.
The structure of each of the three quadratic forms $A^Q_\chi$, $A^Q_{n\chi}$, $B^Q_\chi$ is described by a \emph{symmetric} $2\times2$ matrix, say $q_{ij}$.
Let us first mention that all the matrix elements $q_{ij}$ happen to be positive (which does not, however, imply the positive-definite character of the corresponding quadratic form).
By considering the (orthogonal)  eigendirections and the eigenvalues of $q_{ij}$, we see that, in the case of a form of the type
\be
Q(\cu,\cd) = q_{ij} \cicj,
 \ee
there must be an angle $\phi \in [-\frac{\pi}{2},\frac{\pi}{2})$ such that 
\be
\label{eq:eigdec}
Q = \lambda_1 \lb \cu \cos \phi + \cd \sin \phi \rb^2 + \lambda_2 \lb - \cu \sin \phi + \cd \cos \phi \rb^2
\ee
(and analogously for a form of the type $q_{ij} \nci \ncj$).
Here, for definiteness, $\lambda_1$ denotes the larger eigenvalue, i.e. $\lambda_1 \geq \lambda_2$.
When $\nu = 1/4$, because of the symmetry under exchange of the spins $\cu$ and $\cd$, the only allowed combinations are $\cos \phi = \pm \sin \phi$, thus $\phi(\nu = 1/4) = \pm \pi/4$ in the interval $[-\pi/2,\pi/2)$.
By contrast, the behavior of $\phi$ in the test-mass limit $\nu \to 0$ does not follow a general rule.

\begin{widetext}
\begin{center}
\begin{figure}[h!]
	\includegraphics[scale=0.75]{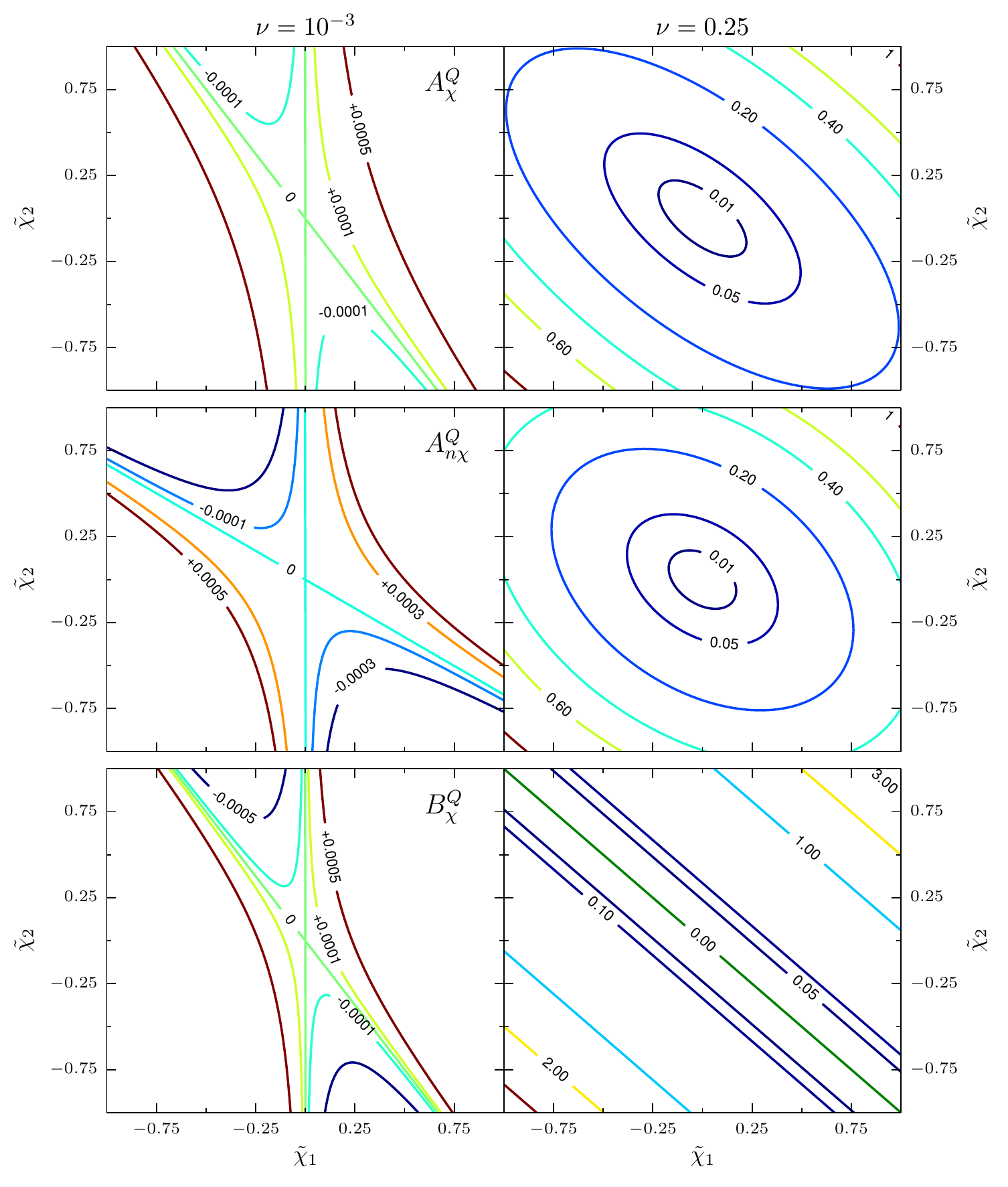}
	\caption{Contour plots of $A^Q_{\chi}$, $A^Q_{n \chi}$ and $B^Q_{\chi}$, each quadratic form corresponding to a row. 
	The two columns correspond to the values $\nu = 10^{-3}$ and $\nu = 0.25$ for which the forms are evaluated. 
	In the case of  $A^Q_{\chi}$ and $B^Q_{\chi}$, aligned or anti-aligned spins are assumed, and the scalar parameters $\tilde{\chi}_i$ have to be interpreted as $\tilde{\chi}_i \equiv \pm |\bm{\chi}_i|$, with $\tilde{\chi}_1 \tilde{\chi}_2 = \cucd$. 
	On the other hand, $\tilde{\chi}_i \equiv \nci $ in the contour plots of $A^Q_{n \chi}$.
	The figures appear to be inclined with respect to a configuration symmetric under reflection of the coordinate axes.
	The measure of such a rotation (in the anti-clockwise direction) is nothing but the angle $\phi$ introduced in Eq.~(\ref{eq:eigdec}) and plotted in Figure~\ref{fig:QuadEig}.   }
	\label{fig:Qlev}
\end{figure}
\end{center}
\end{widetext}

As shown in Figure~\ref{fig:QuadEig}, the eigenvalues $\lambda_1$, $\lambda_2$ of the EOB quadratic forms $A^Q_{\chi}$, $A^Q_{n \chi}$ (and therefore the forms themselves) are positive in most of the range of interest.
For sufficiently small $\nu$, the smaller eigenvalues $\lambda_2$ are negative, and the forms are indefinite.
On the other hand, for larger values of $\nu$, $A^Q_{\chi}$ and $A^Q_{n \chi}$ are both positive definite.

More specifically, the eigenvalues of $A^Q_{\chi}$ are given by
\be
\lambda_{1,2} = \frac{\nu}{2} \lb 3 - \nu \pm \sqrt{13 - 40\nu + \nu^2} \rb,
\ee
with $\lambda_2$ crossing zero at $\nu_0 = 2/17 \approx 0.12$, which corresponds to a mass ratio $m_1/m_2\approx 6.34$.
For circular, equatorial orbits, $\nu > \nu_0$ implies that the new NLO spin-spin terms are always repulsive. 
By contrast, for $\nu < \nu_0$ there are special configurations of the spins where their effect is slightly attractive.

The smallest eigenvalue of $A^Q_{n \chi}$ crosses zero when $\nu = (13 -\sqrt{145})/8 \approx 0.12$.
By contrast with $A^Q_\chi$ and $A^Q_{n\chi}$, $B^Q_{\chi}$ is never positive definite.
However, its largest eigenvalue is always positive, and, most of the time, much larger than $\lambda_2$.
As we shall see later, this implies that $B^Q_\chi$ is positive for most spin configurations.
Note also that $B^Q_\chi$ becomes degenerate ($\lambda_2 =0$) exactly in the case of equal masses ($\nu = 1/4$).

In the two-dimensional parameter space measuring either the projected spins $\nci$, or the algebraic magnitudes of two parallel spins $\cu \parallel \cd$, the contour lines of $Q$ define ellipses, hyperbolas or straight lines, depending on whether $\lambda_2$ is positive, negative or equal to zero, respectively.
A graphical visualization of them is given in Figure~\ref{fig:Qlev}.

The eigenvalue decomposition (\ref{eq:eigdec}) does not provide a direct handle on the extremal points of the quadratic forms.
In order to investigate them, one must resort to other arguments.
Since all coefficients in Eqs.~(\ref{eq:A^Qc})-(\ref{eq:B^Q}) are positive for every $\nu \in (0,1/4]$, it is clear that the global maxima $Q^\text{max}(\nu)$ are reached when $\cu^2 = \cd^2 = \cucd = 1$, or $\ncu = \ncd  = 1$, respectively.

For investigating the minima, let us rewrite 

\be
Q(\cu,\cd) = q_{11} \lb \cu + \frac{q_{12}}{q_{11}} \cd  \rb^2 + \lb q_{22} - \frac{q_{12}^2}{q_{11}}\rb \cd^2.
\ee
If $\lambda_2 <0$, then also $\lb q_{22} - q_{12}^2/q_{11}\rb <0$.
In this case, provided that $q_{12}/q_{11} \leq 1$ (which is indeed true for all quadratic forms (\ref{eq:A^Qc})-(\ref{eq:B^Q})), the global minimum $Q^\text{min}(\nu)$ is reached for the anti-aligned configuration

\begin{align}
\label{eq:antialmin}
\cu = & - \frac{q_{12}}{q_{11}}\cd \text{ , and  } \cd^2 =  1.
\end{align}

Otherwise, if $\lambda_2 \geq 0$, the minimum is met in the trivial case  $\cu = \cd = 0$.
Analogous spin configurations, obtained substituting $\bm{\chi}_i$ with $\nci$ in Eq.~(\ref{eq:antialmin}), define the minima of the forms of the type  $q_{ij} \nci \ncj$.
As a consequence, the extremal values of $B^Q_{\chi}$ and of $B^Q_{n \chi}$ coincide.

\begin{figure}
	\centering
	\includegraphics{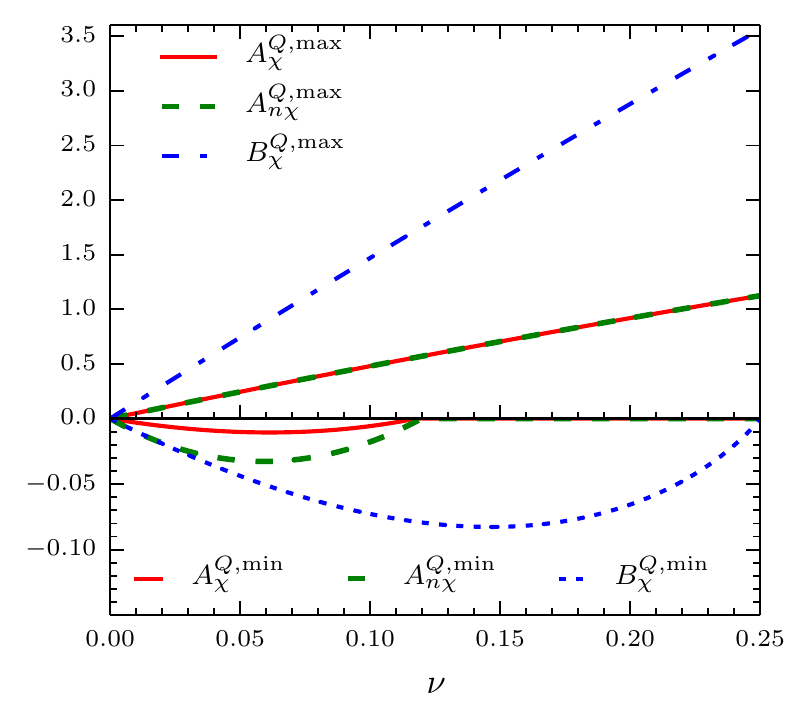}
	\caption{The curves $Q^\text{max}(\nu)$ and $Q^\text{min}(\nu)$ are plotted for the quadratic forms $A^Q_{\chi}$, $A^Q_{n \chi}$ and $B^Q_{\chi}$.
			The region between the two curves represents all possible values that can be taken by the corresponding quadratic form.} 
	\label{fig:QuadMinMax}
\end{figure}

Figure~\ref{fig:QuadMinMax} provides a complete information about the range of values that can be taken by each quadratic form.
Let us remark, in passing, a peculiar feature: although the coefficients of $A^Q_{\chi}$ and of $A^Q_{n \chi}$ could have seemed to be unrelated, they satisfy the identity 

\be
\sum_{ij} a_{ij}^\chi = \sum_{ij} a_{ij}^{n \chi} = \lb 5 - 2\nu \rb \nu.
\ee
Consequently, as is visible on the figure, the maximal curves $A_\chi^{Q \text{,max}}(\nu)$ and $A_{n \chi}^{Q\text{,max}}(\nu)$ are exactly the same.
Among the whole range of $\nu$, their overall maximum is given by $A_\chi^{Q\text{,max}}(1/4) = A_{n \chi}^{Q\text{,max}}(1/4)= 9/8$. 
The overall minimum of $A^Q_{\chi}$ is approximately equal to $-0.011$ and is reached at $\nu \approx 0.061$, while for $A^Q_{n\chi}$ it is reached at $\nu \approx 0.059$ and is nearly equal to $-0.033$.
Moreover, $B_{\chi}^{Q \text{,max}}(1/4) = 57/16$, while the overall minimum $B_{\chi}^{Q \text{,min}} \approx-0.083$ corresponds to $\nu \approx 0.146$.

An order-of-magnitude estimate of the maximal change introduced in $A^\text{eq}$ by $A^Q_{\chi}$ (see Eq.~(\ref{eq:Aeqnew})) can be made by setting $r_c \sim2$ and $A_\chi^{Q\text{,max}} \sim 0.6$, leading to a deviation of $+0.04$ with respect to the LO term $2/r_c \sim 1$.
By contrast, the change in the special configurations where $A^Q_\chi$ is negative is smaller (in absolute value) than $10^{-3}$ , since in this case $A_{\chi}^{Q \text{,min}} \sim - 1/100$.

\section{The spin-orbit sector and the last stable circular orbit}
\label{sec:LSO}
In this last section, we investigate some predictions of the new EOB Hamiltonian proposed here concerning the characteristics of the last stable circular orbit (LSO), considered for parallel spins, and circular, equatorial orbits.

At first, it is necessary to fix the spin-orbit sector $H_\textrm{so}^\textrm{eff}$, that enters the whole effective Hamiltonian as an additive contribution

\be
\hat{H}^\textrm{eff} =  \hat{H}_\textrm{orb}^\textrm{eff} + \hat{H}_\textrm{so}^\textrm{eff}.
\ee
Several different versions of the EOB spin-orbit effective coupling $\hat{H}_\text{so}^\text{eff}$ have been proposed in the literature \cite{Damour:2001tu,Buonanno:2005xu,Damour:2008qf,Barausse:2009xi,Barausse:2009aa,Nagar:2011fx,Barausse:2011ys,Damour:2014sva}.
Here we shall follow the recent approach \cite{Damour:2014sva}, generalizing it to the general, non-equatorial case.
Explicitly, we take 

\be
\label{eq:HsoNew}
\hat{H}_\textrm{so}^\textrm{eff} =  \frac{1}{r\,r_c^2}\lb 1 + \frac{\Delta \ncz^2}{r^2\,r_c^2} \rb^{-1} g_{S}^\text{eff} \bm{l} \cdot \bm{\chi} + \frac{1}{r_c^3}g_{S^*}^\text{eff} \bm{l} \cdot \bm{\chi}^*.
\ee
\begin{center}
\begin{figure}
	\includegraphics{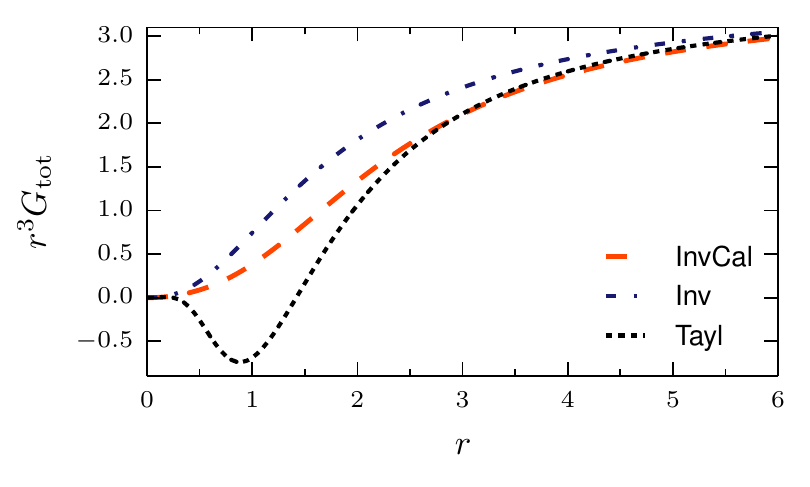}
	\caption{The quantity $r^3 G_\text{tot}$ is plotted against $r$ for circular orbits.
	Equal masses and equal spins $\chi_1 \equiv \chi_2 \equiv 0.65$ are assumed.
	 The curve InvCal corresponds to the model described in Ref.~\cite{Damour:2014sva}, with the NNNLO calibration of $c_3$ described in Ref.~\cite{Nagar:2015xqa}.
	 The curve Inv makes use of the same (inverse) resummation of InvCal, but only includes terms up to NNLO (i.e., it does neither contain the calibrated term $c_3$, nor the two purely Schwarzschild, spinning-particle coefficients that enter into $c_{30}^*$ and  $c_{40}^*$, see Eqs.~(46), (53), (54) in Ref.~\cite{Damour:2014sva}).
	 Finally, Tayl expands the  gyro-gravitomagnetic factors of Inv in a Taylor series.
	 In other words, Tayl is built with the factors $g_{S}^\text{eff}$ and $g_{S^*}^\text{eff}$ as given by Ref.~\cite{Nagar:2011fx}, but with $r_c^{DN14}$ (the centrifugal radius defined in Ref.~\cite{Damour:2014sva}) instead of the Boyer-Lindquist-like radius $r$.
	 The usage of $r_c^{DN14}$ for Tayl has the only goal of allowing a more straightforward comparison against Inv and InvCal.}
	\label{fig:Gtot}
\end{figure}
\end{center}
Here, $\bm{l} \equiv \bm{r} \times \p \equiv \bm{L}/(\mu M)$ is the (dimensionless) rescaled orbital angular momentum, and $\bm{\chi}$ and $\bm{\chi}^*$ are the symmetric spin combinations (\ref{eq:S})-(\ref{eq:S*}), namely

\begin{align}
\bm{\chi} & \equiv \frac{\bm{S}_1 + \bm{S}_2}{(m_1+m_2)^2} = X_1^2 \cu + X_2^2 \cd \\
\bm{\chi}^* & \equiv  \frac{\mdu \bm{S}_1 + \mud \bm{S}_2}{(m_1 + m_2)^2} = \nu \lb \cu + \cd \rb,
\end{align}
while $g_S^\text{eff}$ and $g_{S^*}^\text{eff}$ are two dimensionless gyro-gravitomagnetic factors\footnote{The gyro-gravitomagnetic factors $g_{S}^\text{eff}$ and $g_{S*}^\text{eff}$ used here correspond to $2 \, \hat{G}_{S}$ and $\frac{3}{2}\,\hat{G}_{S^*}$ in Ref.~\cite{Damour:2014sva}.}.
The post-Newtonian expansions of $g_S^\text{eff}$ and $g_{S^*}^\text{eff}$ are fully known up to NNLO order \cite{Damour:2001tu,Damour:2008qf,Nagar:2011fx,Barausse:2011ys}, and one knows both the test-mass limit of $g_{S^*}^\text{eff}$ \cite{Barausse:2009xi} and its first gravitational self-force correction \cite{Bini:2014ica}.

Here, we shall use, as fiducial spin-orbit coupling, the non-resummed, Taylor-expanded NNLO-accurate expansions of $g_S^\text{eff}$ and $g_{S^*}^\text{eff}$ \cite{Nagar:2011fx,Barausse:2011ys}, expressed in the Damour-Jaranowski-Sch\"afer gauge, and (following Ref.~\cite{Damour:2014sva}) using $r_c$ as radial variable.
This means that we use

\begin{widetext}
\begin{align}
	\label{eq:gSeff}
	g_S^\text{eff} =\,& 2 -\frac{27}{8}\nu \np^2 - \frac{5\nu}{8} \frac{1}{r_c} + \frac{5}{8}\nu (1+7\nu) \np^4 
				 +\lb -\frac{21}{2}\nu  + \frac{23}{8}\nu^2 \rb \frac{\np^2}{r_c} - \lb \frac{51}{4}\nu + \frac{\nu^2}{8} \rb \frac{1}{r_c^2} \\
	\label{eq:gSseff}
	g_{S^*}^\text{eff} =\,& \frac{3}{2} -\lb \frac{15}{8} + \frac{9}{4}\nu \rb \np^2 - \lb \frac{9}{8} + \frac{3}{4}\nu \rb\frac{1}{r_c} + \lb \frac{35}{16} + \frac{5}{2}\nu + \frac{45}{16}\nu^2 \rb \np^4 
		 + \lb \frac{69}{16} - \frac{9}{4}\nu + \frac{57}{16}\nu^2 \rb	\frac{\np^2}{r_c} \nonumber \\
	& - \lb \frac{27}{16} + \frac{39}{4}\nu + \frac{3}{16}\nu^2 \rb\frac{1}{r_c^2}.
\end{align}
\end{widetext}

We are aware of the fact that such Taylor-expanded gyro-gravitomagnetic factors have the property of changing sign in the strong-field region, thereby turning the repulsive (for spins parallel to the orbital angular momentum) spin-orbit interaction into an attractive coupling.
In order to avoid this change of sign, Ref.~\cite{Damour:2014sva} used an \emph{inverse} Taylor resummation of the gyro-gravitomagnetic factors (of the type $g_s^\text{eff} = 2/(1+ \frac{\tilde{c}_1}{r_c} + ...)$ etc.).

We compare in Fig~\ref{fig:Gtot} the radial behavior of the total dimensionless effective gyro-gravitomagnetic factors $r^3 G_\text{tot} \equiv r^3\lb \frac{1}{r\,r_c^2} g_S^\text{eff} + \frac{1}{r_c^3}g_{S^*}^\text{eff} \rb$ defined by using either Taylor-expanded $g_S^\text{eff}$, $g_{S^*}^\text{eff}$ or inverse Taylor-expanded ones.
As the main purpose of this subsection is to compare the effect of our new way to incorporate NLO spin-spin coupling to previous suggestions \cite{Balmelli:2013zna, Balmelli:2015lva, Nagar:2015xqa}, it will be convenient for us to use the simple Taylor-expanded prescriptions (\ref{eq:gSeff})-(\ref{eq:gSseff}) because they ensure the existence of an LSO for arbitrary values of the spins.
By contrast, when using inverse-resummed gyro-gravitomagnetic factors the constantly repulsive character of the spin-orbit interaction allows (for large, parallel spins) the sequence of circular orbits to continue existing as the angular momentum decreases, without encountering a loss of stability at some radius.

This is illustrated in Fig~\ref{fig:Heff}  which displays the effective Hamiltonian as a function of radius, for parallel spins equal to $\chi_1 =\chi_2 = 0.65$, and for three different values of the orbital angular momentum: $l =2.7$ (left panel), $l=2.55$ (central panel) and $l=2.4$ (right panel).
 This figure contrasts models which exhibit an LSO for large spins (such as tar14 \cite{Taracchini:2013rva} and models using Taylor-expanded gyro-gravitomagnetic factors, such as our present model, Eq.~(\ref{eq:Aeqnew}), or a version of nag15 \cite{Nagar:2015xqa} in which $g_S^\text{eff}$ and $g_{S^*}^\text{eff}$ are replaced by their Taylor-expanded form) with models that do not, because there exists a continuous sequence of shrinking circular orbits of smaller and smaller radii (such as nag15 \cite{Nagar:2015xqa}).
In particular, it is instructive to compare in Fig~\ref{fig:Heff} the three different versions of the model nag15: (i) the version nag15\_TaylSO (with Taylor-expanded $g_S^\text{eff}$ and $g_{S^*}^\text{eff}$) has an LSO and is quite close to our model (Eq.~(\ref{eq:Aeqnew})); (ii) the version nag15\_NoCal (which differs from \cite{Nagar:2015xqa} by turning off the Numerical-Relativity-calibrated NNLO spin-orbit parameters) displays the strongly repulsive character of the spin-orbit coupling at small radii; and (iii) the original model nag15, which contains extra spin-orbit parameters having the property of reducing (without cancelling) the strongly repulsive character of the spin-orbit coupling.

\begin{widetext}
\begin{center}
\begin{figure}[h!]
	\includegraphics{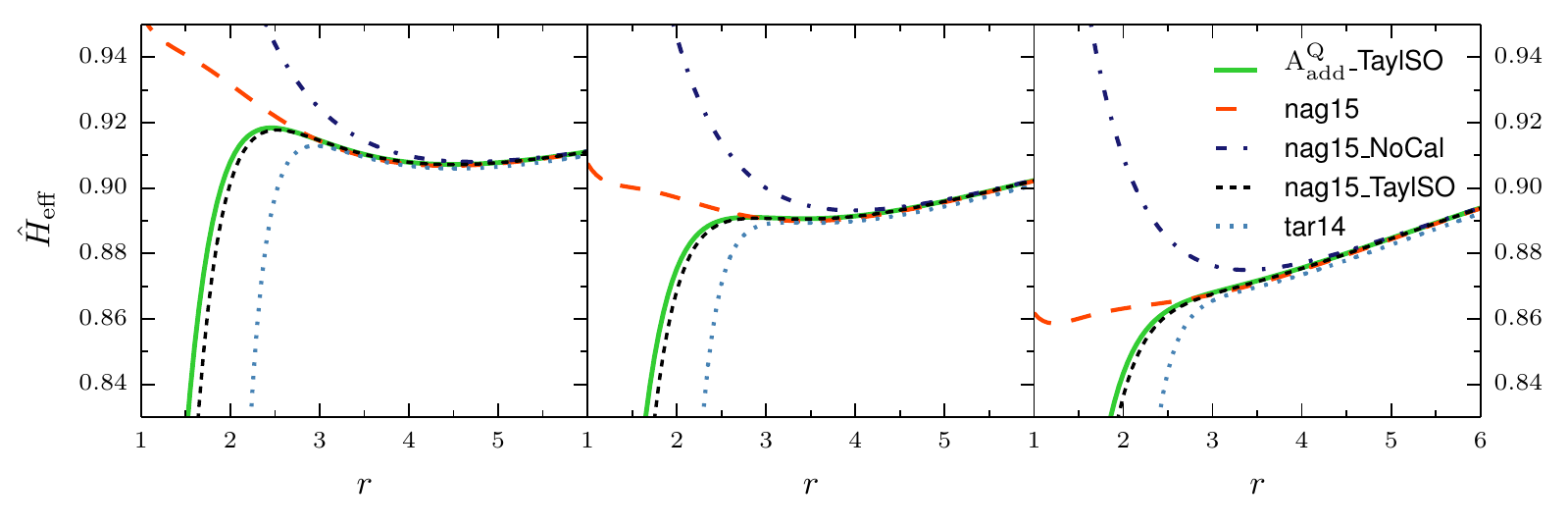}
	\caption{The effective Hamiltonian is plotted as a function of $r$ for circular, equatorial orbits, for parallel spins equal to $\chi_1 =\chi_2 = 0.65$, and for three different values of the orbital angular momentum: $l =2.7$ (left panel), $l=2.55$ (central panel) and $l=2.4$ (right panel).
	The curves tar14 and nag15 denote the calibrated Hamiltonians of Ref.~\cite{Taracchini:2013rva} and of Ref.~\cite{Nagar:2015xqa}, respectively (see the discussion about Fig~\ref{fig:LSO} for some more details); nag15\_NoCal is obtained from nag15 setting to zero the spin-orbit calibration, as well as the two purely Schwarzschild, spinning-particle coefficients that enter into $c_{30}^*$ and  $c_{40}^*$, see Eqs.~(46), (53), (54) in Ref.~\cite{Damour:2014sva}.
	Moreover, nag15\_TaylSO is obtained from nag15\_NoCal by Taylor-expanding its (NNLO) gyro-gravitomagnetic factors.
	Notice that the spin-orbit sector of nag15, nag15\_NoCal and nag15\_TaylSO exactly corresponds to the curves InvCal, Inv and Tayl of Fig~\ref{fig:Gtot}, respectively.
	Finally, $A^Q_\text{add}$\_TaylSO corresponds to the spin-spin model developed in this paper, with a Taylor expanded NNLO spin-orbit sector, and with the same purely orbital terms of nag15.} 
	\label{fig:Heff}
\end{figure}
\end{center}
\end{widetext}
As a consequence, the effective potential of nag15 exhibits (especially for $l=2.4$) a small ``bump'', as if the system would still be trying to develop an LSO.
After this pseudo-LSO, the system rolls down to a further stable minimum, whose existence is ensured by the strong positive spin-orbit barrier.
For sufficiently large spins, the bump ceases to show up, leading therefore to a continuous sequence of circular orbits.
In that case, as for the uncalibrated curve nag15\_NoCal in Fig~\ref{fig:Heff}, the strength of the spin-orbit barrier is such as to completely absorb the region where the LSO would have formed.

The top panels of Fig~\ref{fig:LSO} display a plot of the dimensionless Kerr parameter of the binary system

\be
\chi_J \equiv \frac{1}{\nu} \frac{j_\text{tot}}{\hat{H}_\text{EOB}^2},
\ee
evaluated at the LSO, where

\be
j_\text{tot}\equiv l + \mud \chi_1 + \mdu \chi_2
\ee
is the dimensionless total angular momentum.

\begin{widetext}
\begin{center}
\begin{figure}[h!]
	\includegraphics{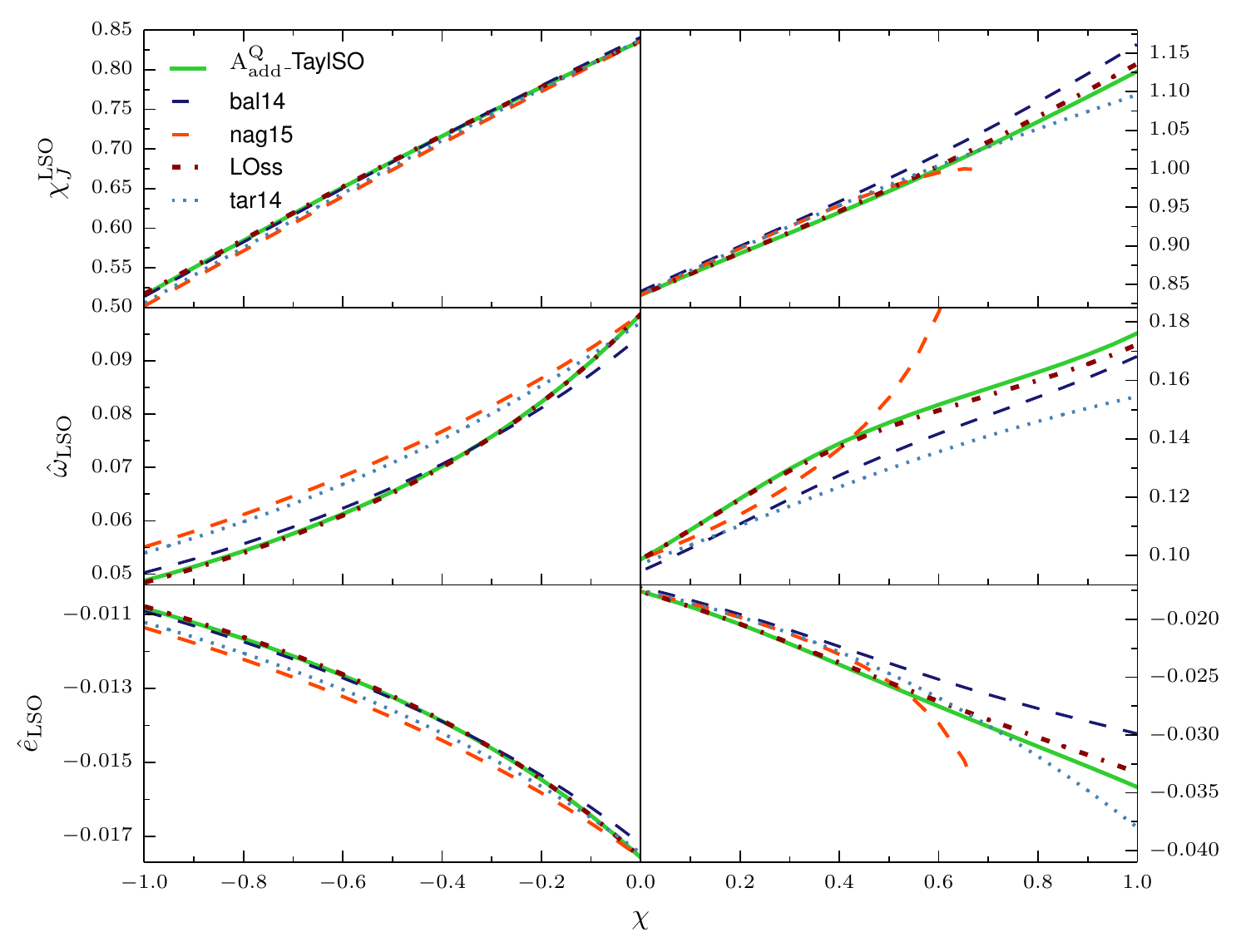}
	\caption{Gauge invariant quantities (top panels: dimensionless total Kerr parameter $\chi_J$; central panels: dimensionless orbital frequency $\hat{\omega}$; bottom panels: dimensionless binding energy $\hat{e}$) at the LSO are plotted as a function of the spin $\chi \equiv \chi_1 \equiv \chi_2$.
	Equal masses are assumed.
	} 
	\label{fig:LSO}
\end{figure}
\end{center}
\end{widetext}

If it were measured after the whole merger-ringdown process, $\chi_J$ would correspond to the dimensionless spin of the final black hole, and would therefore be expected to stay always smaller than one.
At the LSO, however, the system still has to radiate away energy and angular momentum.
It is therefore not worrying to find values $\chi_J^\text{LSO}$ that (slightly) exceed 1 for large spins $\chi \gtrsim 0.6$. 

The central panels plot the dimensionless angular frequency 

\be
\hat{\omega} \equiv \frac{\partial}{\partial l} \hat{H}_\text{EOB},
\ee
and the bottom panels the dimensionless binding energy\footnote{Notice that $\hat{e} =  H_\text{EOB}/M -1$ when expressed in terms of the non-reduced EOB Hamiltonian $H_\text{EOB}$ given by Eq.~(\ref{eq:Heob}).}

\be
\hat{e} = \nu \, \hat{H}_\text{EOB} -1,
\ee
both evaluated at the LSO.
As in Fig~\ref{fig:Heff}, nag15 denotes the calibrated Hamiltonian of Ref.~\cite{Nagar:2015xqa}.
We recall that, in this model, the spin orbit sector is complete up to NNLO and calibrated at the NNNLO level, together with the inclusion of two additional, purely Schwarzschild spinning-particle terms.
Furthermore, the purely orbital coupling is complete at 4PN, and is calibrated at 5PN.
Among all models shown in the figure, this is the only one for which the gyro-gravitomagnetic factors are inversely resummed.
The interruption of the nag15 curves (near $\chi \simeq 0.65$) marks the end of the region where an LSO exists.
Just before reaching that point, a rather strong deviation from the Taylor-spin-orbit curves is clearly visible. 
 
The curves labeled by $A^Q_\text{add}$\_TaylSO denote the spin-spin model developed in this paper, with Taylor expanded, NNLO, $r_c$-dependent gyro-gravitomagnetic factors, while the orbital order is the same as in nag15.
Moreover, LOss represents the curves that are obtained from $A^Q_\text{add}$\_TaylSO by setting $A^Q_\chi$ to zero.
The $A^Q_\text{add}$\_TaylSO and LOss curves are always quite close to each other.
This shows that the difference introduced by the NLO spin-spin coupling is therefore rather small, and by far less important than the effects due to the type of spin-orbit resummation.
The repulsive character of the NLO spin-spin terms, already remarked in Sec~\ref{sec:quad}, is clearly visible on all plots.
Indeed, the total Kerr parameter is smaller than in the LOss, which means that the system radiates away more angular momentum before reaching the end of the inspiral.
Similarly, a larger orbital frequency and binding energy are the signs of a more bound system, and thus imply the existence of an additional repulsive effect preventing the plunge to happen too early.

For completeness, we also show the prediction of the uncalibrated NLO spin-spin Hamiltonian bal14 described in Ref.~\cite{Balmelli:2015lva}.
It is important to remark that bal14 differs from the model of this paper in various aspects, and in particular, it involves a different resummation of both spin-orbit and spin-spin couplings.

Finally, tar14 represents the calibrated model of Ref.~\cite{Taracchini:2013rva}, that encodes the NNLO spin-orbit and LO spin-spin couplings, with a calibration at the NNNLO and NLO level, respectively.
The orbital order is included up to 4PN.
A first aspect to be noticed is the proximity of tar14 with nag15 in the range of negative spins, that can be considered as a qualitative check of the effectiveness of two different calibrations.
For positive spins, the comparison is affected by the different behavior of nag15 for what concerns the LSO.

\begin{table}[t]
  \caption{\label{tab} Dimensionless total Kerr parameter $\chi_J$, orbital frequency $\omega$ and binding energy $\hat{e}$ at the LSO for some values of the spins. 
  Both semi-additive (Add) and factorized (Fact) resummations of $A^Q_\chi$ are shown, together with the case where $A^Q_\chi$ is set to zero (LO).}
    \centering  
  \begin{ruledtabular}  
  \begin{tabular}{llccc}     
    \hline
	& 	$\chi$	&	$\chi_J$		&	$\hat{\omega}$		&	 $\hat{e}$	\\
	\hline
	\hline
LO		&	-1  &   0.5169  &   0.04841 &   -0.01078    \\
Add	 	&	   	&   0.5154  &   0.04877 &   -0.01083    \\
Fact   	&   	& 	0.5148  &   0.04893 &   -0.01085    \\
\hline
LO		&	0.5 &   0.9735  &   0.1441  &   -0.02544    \\
Add	 	&    	&   0.9709  &   0.1456  &   -0.02572    \\
Fact   	&    	&   0.9689  &   0.1472  &   -0.02598    \\
\hline
LO		&	1 	&   1.136   &   0.1723  &   -0.03326    \\
Add	 	&    	&   1.127   &   0.1762  &   -0.03450     \\
Fact   	&    	&   1.118   &   0.1812  &   -0.03587    \\
\hline
\end{tabular}
\end{ruledtabular}  
\end{table}

In Table~\ref{tab} we complement the information contained in Fig~\ref{fig:LSO} by giving a quantitative comparison of the two different resummation options (\ref{eq:AeqnewFact})-(\ref{eq:Aeqnew}) of the $A$ potential, for several values of the spin (namely $-1$, $+0.5$ and $+1$). 
The table confirms the expectation (see Sec~\ref{sec:resum}) that the factorized (Fact) resummation is stronger than the semi-additive (Add) one.
For example, for extremal spins, the increase in the angular frequency at the LSO due to $A^Q_\chi$ is $\simeq +2 \%$ for Add, and $\simeq +5\%$ for Fact, while the binding energy increase is $\simeq +4\%$ (in agreement with the order-of-magnitude estimation done in Sec~\ref{sec:quad}) and $\simeq +8 \%$, respectively.

\section{Conclusions}

In this paper, we have proposed a new EOB Hamiltonian for spinning, precessing black hole binaries.
Explicitly, our Hamiltonian is of the form (\ref{eq:Heob})-(\ref{eq:orb+so}), with an orbital part of the effective Hamiltonian obtained by combining Eqs.~(\ref{eq:Horbeff}), (\ref{eq:A^Qc})-(\ref{eq:rccz}), (\ref{eq:AeqnewFact}) (or (\ref{eq:Aeqnew})), (\ref{eq:Bnpnew}), (\ref{eq:A_B}), and a spin-orbit part defined by combining Eqs.~(\ref{eq:HsoNew})-(\ref{eq:gSseff}).
In particular, we have included spin-spin effects at NLO accuracy by quadratic-in-spin modifications of the building blocks $A(\bm{r}, \nu, \bm{a}_1, \bm{a}_2)$, $B_p(\bm{r},\nu, \bm{a}_1, \bm{a}_2)$, $B_{np}(\bm{r},\nu,\bm{a}_1, \bm{a}_2)$ that are present in the Hamiltonian as coefficients of (part of the) momentum-dependent terms.
Our new approach has several simplifying features with respect to previous works.
First, it maintains a momentum dependence of the squared effective orbital Hamiltonian $\lb H^\text{eff}_\text{orb}\rb^2$ which is no more than quadratic (for the spin-spin terms).
Second, we found that it was possible to choose a spin-gauge where the most complicated NLO spin-spin couplings $\propto \pai \paj$ and $\np \nai \paj$ could be absorbed in a simple Kerr-like coupling $ \propto \enpaz^2$, where $\av_0 \equiv \av_1 + \av_2$ (with $\av_1 \equiv \bm{S}_1/m_1$ and $\av_2 = \bm{S}_2/m_2$) denotes the spin combination describing the LO spin-spin coupling in a Kerr way.
This feature should lead to a simple description of the general precessing spin (and precessing orbital angular momentum) dynamics because of the privileged role of the single basic Kerr-like vectorial spin parameter $\av_0 \equiv \av_1 + \av_2$.

A further tuning allowed us to impose a Damour-Jaranowski-Sch\"afer-type gauge, that has the useful property of confining all new spin-spin terms into the radial potential $A(\bm{r}, \nu, \bm{a}_1, \bm{a}_2)$ as soon as the spins are aligned and the orbits circular.
The NLO spin-spin deformation of the above mentioned sectors is then encoded into quadratic-in-spin forms $A^Q_{\chi}$, $A^Q_{n \chi}$, $B^Q_{\chi}$ and $B^Q_{n \chi}$, see Eq.~(\ref{eq:A^Qc})-(\ref{eq:B^Qnc}), which are our main results.
A remarkable fact is that the coefficients of $B^Q_{\chi}$ and of $B^Q_{n \chi}$ are exactly the same.
Therefore, the 25 independent coefficients that define the NLO spin-spin Hamiltonian in ADM coordinates shrink down to only 9 in the EOB description.
A further, minor symmetry property lies in the fact that the \emph{sum} of all the coefficients of $A^Q_{\chi}$ and of $A^Q_{n \chi}$ are equal.
These features correspond to a notable improvement with respect to the model developed in Ref.~\cite{Balmelli:2015lva}, where the momentum structure of spin-dependent terms is by far less simple (for instance, the squared  effective orbital Hamiltonian of Ref.~\cite{Balmelli:2015lva} does not show a polynomial dependence on the momenta, and furthermore no Damour-Jaranowski-Sch\"afer-type gauge could be imposed) and where the number of independent NLO spin-spin coefficients to be inserted in the EOB description amounts to 12.

The quadratic forms we have found here have positive coefficients only.
However, as quadratic forms, they are either indefinite (with a positive eigenvalue and a negative one), degenerate (with one eigenvalue being strictly positive and the other zero) or positive definite, depending on the value of the symmetric mass ratio $\nu$. 
For sufficienly low $\nu$, the smaller eigenvalue is negative, and the form is negative-valued for particular configurations of anti-aligned, or nearly anti-aligned spins.
By contrast, aligned configurations always lead to positive values, that are moreover much larger (by a factor $\sim$ 50-100) than the negative minima.
For what concerns circular, equatorial orbits, one can conclude that the NLO spin-spin effects are repulsive in most cases, apart from very small, attractive effects that only show up for mass ratios $m_1/m_2 \geq 6.34$ and for (nearly) anti-aligned spins.
This repulsive character is clearly visible when comparing the total angular momentum, angular frequency and binding energy at the LSO with the corresponding prediction of the Hamiltonian without the NLO spin-spin inclusion.
We propose two different options for resumming  the quadratic form $A^Q_\chi$, a semi-additive and a factorized one. 
The ultimate choice of the best resummation option can only be done with a systematic comparison against Numerical Relativity simulations.
We expect our new Hamiltonian, once calibrated, to mark a new step towards an accurate description of the coalescence of two precessing, spinning black holes.

\begin{acknowledgments}
S. B. thanks IHES for hospitality during the development of the main part of this work. He is supported by the Swiss National Science Foundation.
\end{acknowledgments}

\appendix
\setcounter{secnumdepth}{0}
\renewcommand{\theequation}{A.\arabic{equation}}
\section{Appendix: On the hidden ``symmetry'' of the NLO spin-spin coupling}
\label{sec:app}

We have seen in the text that the (effective) EOB Hamiltonian was exhibiting six remarkable cancellations and/or coincidences among the spin-quadratic forms describing the NLO spin-spin coupling.
Namely, in our preferred gauge-fixing, these six remarkable ``symmetries'' amounted to the equations ($i,j=1,2$)
\be
\label{eq:symm}
b^{np,n\chi}_{ij} \equiv 0, \quad \quad  b^{np,\chi}_{ij} \equiv b^{p,n\chi}_{ij}.
\ee
These 6 symmetries, together with the appropriate use of the 10 NLO gauge parameters contained in $\hat{G}_\text{ss}^\text{NLO}$, has allowed us to end up with a final EOB Hamiltonian containing only 9 different coefficients to describe the NLO spin-spin sector, when starting from the ADM spin-spin Hamiltonian which contained 25 different NLO spin-spin coefficients.
In this Appendix, we trace the origin of these six symmetries in the original ADM Hamiltonian.
Let us denote the momentum-dependent part of a NLO spin-spin Hamiltonian as
\begin{align}
\hat{H}_\text{ss}^{\text{NLO}}|_{p\text{-dep}} = 
&\frac{1}{r^3} \Big[ 
	\lb c_1^{ij} \p^2 + c_2^{ij} \np^2\rb \cicj \nonumber \\
&	\lb c_3^{ij} \p^2 + c_4^{ij} \np^2\rb \nci \ncj \nonumber \\
&	+ c_5^{ij} \pci \pcj \nonumber \\
&	+ c_6^{ij} \np \pci \ncj  
	\Big].
\end{align}
Because of the variation structure described by Eq.~(\ref{eq:deltaH}), under a canonical transformation 
\be
		\hat{\tilde{H}}_\text{ss}^{\text{NLO}} = \hat{H}_\text{ss}^{\text{NLO(ADM)} } + \big\{ \hat{G}_\text{ss}^\text{NLO}, \hat{H}_\text{N} \big\},
\ee
one can easily check that the combinations $3c_1^{ij} + c_2^{ij}$, $5c_3^{ij} + c_4^{ij}$ and $-2c_3^{ij} +3c_5^{ij} +c_6^{(ij)}$ are gauge invariant.
We can further check (from the explicit expressions of the ADM coefficients) that the 6 following gauge-invariant combinations of coefficients happen to vanish:
\begin{subequations}
\label{eq:gginv}
\begin{align}
\label{eq:gginv2}
3c_1^{ij} + c_2^{ij} + c_3^{ij} + \frac{c_4^{ij}}{5} &= 0 \\
\label{eq:gginv3}
3c_1^{ij} + c_2^{ij} + c_3^{ij} -\frac{3}{2}c_5^{ij} - \frac{c_6^{(ij)}}{2} &= 0.
\end{align}
\end{subequations}
One can consider that the six identities (\ref{eq:gginv}) constitute the hidden origin of the six (more manifest) relations (\ref{eq:symm}) found in their EOB transcription.
In that sense, one can say that the EOB formulation is useful in revealing, and making manifest, symmetries that existed, in a hidden way, as 6 relations between the 25 original ADM coefficients.
So that, finally, there is, as expected, a conservation of linearly independent NLO spin-spin coefficients, with $9 = 25 - 10\text{(gauge)} - 6\text{(relations)}$.

\bibliography{refs.bib}

\begin{thebibliography}{35}
\expandafter\ifx\csname natexlab\endcsname\relax\def\natexlab#1{#1}\fi
\expandafter\ifx\csname bibnamefont\endcsname\relax
  \def\bibnamefont#1{#1}\fi
\expandafter\ifx\csname bibfnamefont\endcsname\relax
  \def\bibfnamefont#1{#1}\fi
\expandafter\ifx\csname citenamefont\endcsname\relax
  \def\citenamefont#1{#1}\fi
\expandafter\ifx\csname url\endcsname\relax
  \def\url#1{\texttt{#1}}\fi
\expandafter\ifx\csname urlprefix\endcsname\relax\def\urlprefix{URL }\fi
\providecommand{\bibinfo}[2]{#2}
\providecommand{\eprint}[2][]{\url{#2}}

\bibitem[{vir()}]{virgo}
\bibinfo{howpublished}{\url{http://www.ego-gw.it/}}.

\bibitem[{\citenamefont{Aasi et~al.}(2015)}]{TheLIGOScientific:2014jea}
\bibinfo{author}{\bibfnamefont{J.}~\bibnamefont{Aasi}} \bibnamefont{et~al.}
  (\bibinfo{collaboration}{LIGO Scientific}), \bibinfo{journal}{Class. Quant.
  Grav.} \textbf{\bibinfo{volume}{32}}, \bibinfo{pages}{074001}
  (\bibinfo{year}{2015}), \eprint{1411.4547}.

\bibitem[{\citenamefont{Faye et~al.}(2006)\citenamefont{Faye, Blanchet, and
  Buonanno}}]{Faye:2006gx}
\bibinfo{author}{\bibfnamefont{G.}~\bibnamefont{Faye}},
  \bibinfo{author}{\bibfnamefont{L.}~\bibnamefont{Blanchet}}, \bibnamefont{and}
  \bibinfo{author}{\bibfnamefont{A.}~\bibnamefont{Buonanno}},
  \bibinfo{journal}{Phys.Rev.} \textbf{\bibinfo{volume}{D74}},
  \bibinfo{pages}{104033} (\bibinfo{year}{2006}), \eprint{gr-qc/0605139}.

\bibitem[{\citenamefont{Blanchet et~al.}(2006)\citenamefont{Blanchet, Buonanno,
  and Faye}}]{Blanchet:2006gy}
\bibinfo{author}{\bibfnamefont{L.}~\bibnamefont{Blanchet}},
  \bibinfo{author}{\bibfnamefont{A.}~\bibnamefont{Buonanno}}, \bibnamefont{and}
  \bibinfo{author}{\bibfnamefont{G.}~\bibnamefont{Faye}},
  \bibinfo{journal}{Phys.Rev.} \textbf{\bibinfo{volume}{D74}},
  \bibinfo{pages}{104034} (\bibinfo{year}{2006}), \eprint{gr-qc/0605140}.

\bibitem[{\citenamefont{Damour et~al.}(2008{\natexlab{a}})\citenamefont{Damour,
  Jaranowski, and Schaefer}}]{Damour:2007nc}
\bibinfo{author}{\bibfnamefont{T.}~\bibnamefont{Damour}},
  \bibinfo{author}{\bibfnamefont{P.}~\bibnamefont{Jaranowski}},
  \bibnamefont{and} \bibinfo{author}{\bibfnamefont{G.}~\bibnamefont{Schaefer}},
  \bibinfo{journal}{Phys. Rev.} \textbf{\bibinfo{volume}{D77}},
  \bibinfo{pages}{064032} (\bibinfo{year}{2008}{\natexlab{a}}).

\bibitem[{\citenamefont{Steinhoff
  et~al.}(2008{\natexlab{a}})\citenamefont{Steinhoff, Schaefer, and
  Hergt}}]{Steinhoff:2008zr}
\bibinfo{author}{\bibfnamefont{J.}~\bibnamefont{Steinhoff}},
  \bibinfo{author}{\bibfnamefont{G.}~\bibnamefont{Schaefer}}, \bibnamefont{and}
  \bibinfo{author}{\bibfnamefont{S.}~\bibnamefont{Hergt}},
  \bibinfo{journal}{Phys. Rev.} \textbf{\bibinfo{volume}{D77}},
  \bibinfo{pages}{104018} (\bibinfo{year}{2008}{\natexlab{a}}),
  \eprint{0805.3136}.

\bibitem[{\citenamefont{Steinhoff and Schaefer}(2009)}]{Steinhoff:2009ei}
\bibinfo{author}{\bibfnamefont{J.}~\bibnamefont{Steinhoff}} \bibnamefont{and}
  \bibinfo{author}{\bibfnamefont{G.}~\bibnamefont{Schaefer}},
  \bibinfo{journal}{Europhys. Lett.} \textbf{\bibinfo{volume}{87}},
  \bibinfo{pages}{50004} (\bibinfo{year}{2009}), \eprint{0907.1967}.

\bibitem[{\citenamefont{Hartung and Steinhoff}(2011)}]{Hartung:2011te}
\bibinfo{author}{\bibfnamefont{J.}~\bibnamefont{Hartung}} \bibnamefont{and}
  \bibinfo{author}{\bibfnamefont{J.}~\bibnamefont{Steinhoff}},
  \bibinfo{journal}{Annalen Phys.} \textbf{\bibinfo{volume}{523}},
  \bibinfo{pages}{783} (\bibinfo{year}{2011}), \eprint{1104.3079}.

\bibitem[{\citenamefont{Hergt and Schaefer}(2008)}]{Hergt:2008jn}
\bibinfo{author}{\bibfnamefont{S.}~\bibnamefont{Hergt}} \bibnamefont{and}
  \bibinfo{author}{\bibfnamefont{G.}~\bibnamefont{Schaefer}},
  \bibinfo{journal}{Phys. Rev.} \textbf{\bibinfo{volume}{D78}},
  \bibinfo{pages}{124004} (\bibinfo{year}{2008}), \eprint{0809.2208}.

\bibitem[{\citenamefont{Steinhoff
  et~al.}(2008{\natexlab{b}})\citenamefont{Steinhoff, Hergt, and
  Schaefer}}]{Steinhoff:2008ji}
\bibinfo{author}{\bibfnamefont{J.}~\bibnamefont{Steinhoff}},
  \bibinfo{author}{\bibfnamefont{S.}~\bibnamefont{Hergt}}, \bibnamefont{and}
  \bibinfo{author}{\bibfnamefont{G.}~\bibnamefont{Schaefer}},
  \bibinfo{journal}{Phys.Rev.} \textbf{\bibinfo{volume}{D78}},
  \bibinfo{pages}{101503} (\bibinfo{year}{2008}{\natexlab{b}}),
  \eprint{0809.2200}.

\bibitem[{\citenamefont{Steinhoff
  et~al.}(2008{\natexlab{c}})\citenamefont{Steinhoff, Hergt, and
  Schaefer}}]{Steinhoff:2007mb}
\bibinfo{author}{\bibfnamefont{J.}~\bibnamefont{Steinhoff}},
  \bibinfo{author}{\bibfnamefont{S.}~\bibnamefont{Hergt}}, \bibnamefont{and}
  \bibinfo{author}{\bibfnamefont{G.}~\bibnamefont{Schaefer}},
  \bibinfo{journal}{Phys.Rev.} \textbf{\bibinfo{volume}{D77}},
  \bibinfo{pages}{081501} (\bibinfo{year}{2008}{\natexlab{c}}),
  \eprint{0712.1716}.

\bibitem[{\citenamefont{Goldberger and Rothstein}(2006)}]{Goldberger:2004jt}
\bibinfo{author}{\bibfnamefont{W.~D.} \bibnamefont{Goldberger}}
  \bibnamefont{and} \bibinfo{author}{\bibfnamefont{I.~Z.}
  \bibnamefont{Rothstein}}, \bibinfo{journal}{Phys. Rev.}
  \textbf{\bibinfo{volume}{D73}}, \bibinfo{pages}{104029}
  (\bibinfo{year}{2006}), \eprint{hep-th/0409156}.

\bibitem[{\citenamefont{Levi and Steinhoff}(2015{\natexlab{a}})}]{Levi:2015msa}
\bibinfo{author}{\bibfnamefont{M.}~\bibnamefont{Levi}} \bibnamefont{and}
  \bibinfo{author}{\bibfnamefont{J.}~\bibnamefont{Steinhoff}}
  (\bibinfo{year}{2015}{\natexlab{a}}), \eprint{1501.04956}.

\bibitem[{\citenamefont{Levi and Steinhoff}(2015{\natexlab{b}})}]{Levi:2015ixa}
\bibinfo{author}{\bibfnamefont{M.}~\bibnamefont{Levi}} \bibnamefont{and}
  \bibinfo{author}{\bibfnamefont{J.}~\bibnamefont{Steinhoff}}
  (\bibinfo{year}{2015}{\natexlab{b}}), \eprint{1506.05794}.

\bibitem[{\citenamefont{Buonanno and Damour}(1999)}]{Buonanno:1998gg}
\bibinfo{author}{\bibfnamefont{A.}~\bibnamefont{Buonanno}} \bibnamefont{and}
  \bibinfo{author}{\bibfnamefont{T.}~\bibnamefont{Damour}},
  \bibinfo{journal}{Phys. Rev.} \textbf{\bibinfo{volume}{D59}},
  \bibinfo{pages}{084006} (\bibinfo{year}{1999}).

\bibitem[{\citenamefont{Buonanno and Damour}(2000)}]{Buonanno:2000ef}
\bibinfo{author}{\bibfnamefont{A.}~\bibnamefont{Buonanno}} \bibnamefont{and}
  \bibinfo{author}{\bibfnamefont{T.}~\bibnamefont{Damour}},
  \bibinfo{journal}{Phys. Rev.} \textbf{\bibinfo{volume}{D62}},
  \bibinfo{pages}{064015} (\bibinfo{year}{2000}).

\bibitem[{\citenamefont{Damour et~al.}(2000)\citenamefont{Damour, Jaranowski,
  and Schaefer}}]{Damour:2000we}
\bibinfo{author}{\bibfnamefont{T.}~\bibnamefont{Damour}},
  \bibinfo{author}{\bibfnamefont{P.}~\bibnamefont{Jaranowski}},
  \bibnamefont{and} \bibinfo{author}{\bibfnamefont{G.}~\bibnamefont{Schaefer}},
  \bibinfo{journal}{Phys. Rev.} \textbf{\bibinfo{volume}{D62}},
  \bibinfo{pages}{084011} (\bibinfo{year}{2000}).

\bibitem[{\citenamefont{Damour}(2001)}]{Damour:2001tu}
\bibinfo{author}{\bibfnamefont{T.}~\bibnamefont{Damour}},
  \bibinfo{journal}{Phys. Rev.} \textbf{\bibinfo{volume}{D64}},
  \bibinfo{pages}{124013} (\bibinfo{year}{2001}).

\bibitem[{\citenamefont{Damour et~al.}(2009)\citenamefont{Damour, Iyer, and
  Nagar}}]{Damour:2008gu}
\bibinfo{author}{\bibfnamefont{T.}~\bibnamefont{Damour}},
  \bibinfo{author}{\bibfnamefont{B.~R.} \bibnamefont{Iyer}}, \bibnamefont{and}
  \bibinfo{author}{\bibfnamefont{A.}~\bibnamefont{Nagar}},
  \bibinfo{journal}{Phys. Rev.} \textbf{\bibinfo{volume}{D79}},
  \bibinfo{pages}{064004} (\bibinfo{year}{2009}).

\bibitem[{\citenamefont{Buonanno et~al.}(2006)\citenamefont{Buonanno, Chen, and
  Damour}}]{Buonanno:2005xu}
\bibinfo{author}{\bibfnamefont{A.}~\bibnamefont{Buonanno}},
  \bibinfo{author}{\bibfnamefont{Y.}~\bibnamefont{Chen}}, \bibnamefont{and}
  \bibinfo{author}{\bibfnamefont{T.}~\bibnamefont{Damour}},
  \bibinfo{journal}{Phys. Rev.} \textbf{\bibinfo{volume}{D74}},
  \bibinfo{pages}{104005} (\bibinfo{year}{2006}), \eprint{gr-qc/0508067}.

\bibitem[{\citenamefont{Damour et~al.}(2008{\natexlab{b}})\citenamefont{Damour,
  Jaranowski, and Schaefer}}]{Damour:2008qf}
\bibinfo{author}{\bibfnamefont{T.}~\bibnamefont{Damour}},
  \bibinfo{author}{\bibfnamefont{P.}~\bibnamefont{Jaranowski}},
  \bibnamefont{and} \bibinfo{author}{\bibfnamefont{G.}~\bibnamefont{Schaefer}},
  \bibinfo{journal}{Phys.Rev.} \textbf{\bibinfo{volume}{D78}},
  \bibinfo{pages}{024009} (\bibinfo{year}{2008}{\natexlab{b}}),
  \eprint{0803.0915}.

\bibitem[{\citenamefont{Barausse et~al.}(2009)\citenamefont{Barausse, Racine,
  and Buonanno}}]{Barausse:2009aa}
\bibinfo{author}{\bibfnamefont{E.}~\bibnamefont{Barausse}},
  \bibinfo{author}{\bibfnamefont{E.}~\bibnamefont{Racine}}, \bibnamefont{and}
  \bibinfo{author}{\bibfnamefont{A.}~\bibnamefont{Buonanno}},
  \bibinfo{journal}{Phys. Rev.} \textbf{\bibinfo{volume}{D80}},
  \bibinfo{pages}{104025} (\bibinfo{year}{2009}), \eprint{0907.4745}.

\bibitem[{\citenamefont{Barausse and Buonanno}(2010)}]{Barausse:2009xi}
\bibinfo{author}{\bibfnamefont{E.}~\bibnamefont{Barausse}} \bibnamefont{and}
  \bibinfo{author}{\bibfnamefont{A.}~\bibnamefont{Buonanno}},
  \bibinfo{journal}{Phys.Rev.} \textbf{\bibinfo{volume}{D81}},
  \bibinfo{pages}{084024} (\bibinfo{year}{2010}), \eprint{0912.3517}.

\bibitem[{\citenamefont{Pan et~al.}(2011)\citenamefont{Pan, Buonanno, Fujita,
  Racine, and Tagoshi}}]{Pan:2010hz}
\bibinfo{author}{\bibfnamefont{Y.}~\bibnamefont{Pan}},
  \bibinfo{author}{\bibfnamefont{A.}~\bibnamefont{Buonanno}},
  \bibinfo{author}{\bibfnamefont{R.}~\bibnamefont{Fujita}},
  \bibinfo{author}{\bibfnamefont{E.}~\bibnamefont{Racine}}, \bibnamefont{and}
  \bibinfo{author}{\bibfnamefont{H.}~\bibnamefont{Tagoshi}},
  \bibinfo{journal}{Phys.Rev.} \textbf{\bibinfo{volume}{D83}},
  \bibinfo{pages}{064003} (\bibinfo{year}{2011}), \eprint{1006.0431}.

\bibitem[{\citenamefont{Nagar}(2011)}]{Nagar:2011fx}
\bibinfo{author}{\bibfnamefont{A.}~\bibnamefont{Nagar}},
  \bibinfo{journal}{Phys.Rev.} \textbf{\bibinfo{volume}{D84}},
  \bibinfo{pages}{084028} (\bibinfo{year}{2011}), \eprint{1106.4349}.

\bibitem[{\citenamefont{Barausse and Buonanno}(2011)}]{Barausse:2011ys}
\bibinfo{author}{\bibfnamefont{E.}~\bibnamefont{Barausse}} \bibnamefont{and}
  \bibinfo{author}{\bibfnamefont{A.}~\bibnamefont{Buonanno}},
  \bibinfo{journal}{Phys.Rev.} \textbf{\bibinfo{volume}{D84}},
  \bibinfo{pages}{104027} (\bibinfo{year}{2011}), \eprint{1107.2904}.

\bibitem[{\citenamefont{Pan et~al.}(2014)\citenamefont{Pan, Buonanno,
  Taracchini, Kidder, Mrou\'e, Pfeiffer, Scheel, and Szil\'agyi}}]{Pan:2013rra}
\bibinfo{author}{\bibfnamefont{Y.}~\bibnamefont{Pan}},
  \bibinfo{author}{\bibfnamefont{A.}~\bibnamefont{Buonanno}},
  \bibinfo{author}{\bibfnamefont{A.}~\bibnamefont{Taracchini}},
  \bibinfo{author}{\bibfnamefont{L.~E.} \bibnamefont{Kidder}},
  \bibinfo{author}{\bibfnamefont{A.~H.} \bibnamefont{Mrou\'e}},
  \bibinfo{author}{\bibfnamefont{H.~P.} \bibnamefont{Pfeiffer}},
  \bibinfo{author}{\bibfnamefont{M.~A.} \bibnamefont{Scheel}},
  \bibnamefont{and}
  \bibinfo{author}{\bibfnamefont{B.}~\bibnamefont{Szil\'agyi}},
  \bibinfo{journal}{Phys. Rev. D} \textbf{\bibinfo{volume}{89}},
  \bibinfo{pages}{084006} (\bibinfo{year}{2014}).

\bibitem[{\citenamefont{Balmelli and Jetzer}(2013)}]{Balmelli:2013zna}
\bibinfo{author}{\bibfnamefont{S.}~\bibnamefont{Balmelli}} \bibnamefont{and}
  \bibinfo{author}{\bibfnamefont{P.}~\bibnamefont{Jetzer}},
  \bibinfo{journal}{Phys.Rev.} \textbf{\bibinfo{volume}{D87}},
  \bibinfo{pages}{124036} (\bibinfo{year}{2013}), \eprint{1305.5674}.

\bibitem[{\citenamefont{Balmelli and Jetzer}(2014)}]{Balmelli:2013err}
\bibinfo{author}{\bibfnamefont{S.}~\bibnamefont{Balmelli}} \bibnamefont{and}
  \bibinfo{author}{\bibfnamefont{P.}~\bibnamefont{Jetzer}},
  \bibinfo{journal}{Phys. Rev. D} \textbf{\bibinfo{volume}{90}},
  \bibinfo{pages}{089905(E)} (\bibinfo{year}{2014}).

\bibitem[{\citenamefont{Balmelli and Jetzer}(2015)}]{Balmelli:2015lva}
\bibinfo{author}{\bibfnamefont{S.}~\bibnamefont{Balmelli}} \bibnamefont{and}
  \bibinfo{author}{\bibfnamefont{P.}~\bibnamefont{Jetzer}},
  \bibinfo{journal}{Phys.Rev.} \textbf{\bibinfo{volume}{D91}},
  \bibinfo{pages}{064011} (\bibinfo{year}{2015}), \eprint{1502.01343}.

\bibitem[{\citenamefont{Damour and Nagar}(2014)}]{Damour:2014sva}
\bibinfo{author}{\bibfnamefont{T.}~\bibnamefont{Damour}} \bibnamefont{and}
  \bibinfo{author}{\bibfnamefont{A.}~\bibnamefont{Nagar}},
  \bibinfo{journal}{Phys.Rev.} \textbf{\bibinfo{volume}{D90}},
  \bibinfo{pages}{044018} (\bibinfo{year}{2014}), \eprint{1406.6913}.

\bibitem[{\citenamefont{Bini and Damour}(2013)}]{Bini:2013zaa}
\bibinfo{author}{\bibfnamefont{D.}~\bibnamefont{Bini}} \bibnamefont{and}
  \bibinfo{author}{\bibfnamefont{T.}~\bibnamefont{Damour}},
  \bibinfo{journal}{Phys.Rev.} \textbf{\bibinfo{volume}{D87}},
  \bibinfo{pages}{121501} (\bibinfo{year}{2013}), \eprint{1305.4884}.

\bibitem[{\citenamefont{Nagar et~al.}(2015)\citenamefont{Nagar, Damour,
  Reisswig, and Pollney}}]{Nagar:2015xqa}
\bibinfo{author}{\bibfnamefont{A.}~\bibnamefont{Nagar}},
  \bibinfo{author}{\bibfnamefont{T.}~\bibnamefont{Damour}},
  \bibinfo{author}{\bibfnamefont{C.}~\bibnamefont{Reisswig}}, \bibnamefont{and}
  \bibinfo{author}{\bibfnamefont{D.}~\bibnamefont{Pollney}}
  (\bibinfo{year}{2015}), \eprint{1506.08457}.

\bibitem[{\citenamefont{Bini and Damour}(2014)}]{Bini:2014ica}
\bibinfo{author}{\bibfnamefont{D.}~\bibnamefont{Bini}} \bibnamefont{and}
  \bibinfo{author}{\bibfnamefont{T.}~\bibnamefont{Damour}},
  \bibinfo{journal}{Phys. Rev.} \textbf{\bibinfo{volume}{D90}},
  \bibinfo{pages}{024039} (\bibinfo{year}{2014}), \eprint{1404.2747}.

\bibitem[{\citenamefont{Taracchini et~al.}(2014)\citenamefont{Taracchini,
  Buonanno, Pan, Hinderer, Boyle et~al.}}]{Taracchini:2013rva}
\bibinfo{author}{\bibfnamefont{A.}~\bibnamefont{Taracchini}},
  \bibinfo{author}{\bibfnamefont{A.}~\bibnamefont{Buonanno}},
  \bibinfo{author}{\bibfnamefont{Y.}~\bibnamefont{Pan}},
  \bibinfo{author}{\bibfnamefont{T.}~\bibnamefont{Hinderer}},
  \bibinfo{author}{\bibfnamefont{M.}~\bibnamefont{Boyle}},
  \bibnamefont{et~al.}, \bibinfo{journal}{Phys.Rev.}
  \textbf{\bibinfo{volume}{D89}}, \bibinfo{pages}{061502}
  (\bibinfo{year}{2014}), \eprint{1311.2544}.

\end{thebibliography}

\bibliographystyle{apsrev}

\end{document}